\documentclass[pdflatex,sn-mathphys-num]{sn-jnl}

\usepackage{graphicx}%
\usepackage{multirow}%
\usepackage{amsmath,amssymb,amsfonts}%
\usepackage{amsthm}%
\usepackage{mathrsfs}%
\usepackage[title]{appendix}%
\usepackage{xcolor}%
\usepackage{textcomp}%
\usepackage{manyfoot}%
\usepackage{booktabs}%
\usepackage{braket}
\usepackage{algorithm}%
\usepackage{algorithmicx}%
\usepackage{algpseudocode}%
\usepackage{listings}%
\usepackage{bm}

\theoremstyle{thmstyleone}%
\theoremstyle{thmstyletwo}%

\theoremstyle{thmstylethree}%

\begin{document}

\title[Article Title]{Emulation of Coherent Absorption of Quantum Light in a Programmable Linear Photonic Circuit}


\author*[1]{\fnm{Govind} \sur{Krishna}}\email{govindk@kth.se}

\author*[1]{\fnm{Jun} \sur{Gao}}\email{junga@kth.se}

\author[1]{\fnm{Sam} \sur{O’Brien}}

\author[1]{\fnm{Rohan} \sur{Yadgirkar}}

\author[2]{\fnm{Venkatesh} \sur{Deenadayalan}}

\author[2]{\fnm{Stefan} \sur{Preble}}

\author[1]{\fnm{Val} \sur{Zwiller}}

\author*[1]{\fnm{Ali} \sur{W. Elshaari}}\email{elshaari@kth.se}

\affil[1]{\orgdiv{Department of Applied Physics}, \orgname{KTH Royal Institute of Technology}, \orgaddress{\street{Albanova University Centre, Roslagstullsbacken 21}, \city{Stockholm}, \postcode{106 91}, \state{Stockholm}, \country{Sweden}}}

\affil[2]{\orgdiv{Microsystems Engineering}, \orgname{Rochester Institute of Technology}, \orgaddress{\street{Street}, \city{Rochester}, \postcode{14623}, \state{New York}, \country{USA}}}


\abstract{Non-Hermitian quantum systems, governed by nonunitary evolution, offer powerful tools for manipulating quantum states through engineered loss. A prime example is coherent absorption, where quantum states undergo phase-dependent partial or complete absorption in a lossy medium. Here, we demonstrate a fully programmable implementation of nonunitary transformations that emulate coherent absorption of quantum light using a programmable integrated linear photonic circuit, with loss introduced via coupling to an ancilla mode [Phys. Rev. X 8, 021017; 2018]. Probing the circuit with a single-photon dual-rail state reveals phase-controlled coherent tunability between perfect transmission and perfect absorption. A two-photon NOON state input, by contrast, exhibits switching between deterministic single-photon and probabilistic two-photon absorption. Across a range of input phases and circuit configurations, we observe nonclassical effects such as anti-coalescence and bunching, along with continuous and coherent tuning of output Fock state probability amplitudes. Classical Fisher information analysis reveals phase sensitivity peaks of 1 for single-photon states and 3.4 for NOON states, the latter exceeding the shot-noise limit of 2 and approaching the Heisenberg limit of 4 for two-photon states. The experiment integrates quantum state generation, programmable photonic circuitry, and photon-number-resolving detection, establishing ancilla-assisted circuits as powerful tools for programmable quantum state engineering, filtering, multiplexed sensing, and nonunitary quantum simulation.}




\maketitle

Non-unitary transformations describe irreversible processes such as loss, gain, or measurement back-action in open systems, in contrast to the reversible unitary dynamics of isolated systems~\cite{Breuer2007}. In quantum photonics, unitary transformations preserve photon number and are typically implemented via multiport interferometer networks~\cite{Reck1994,Clements2016}. Extending linear optics to the non-unitary domain enables modeling dissipation, decoherence, amplification, and generalized quantum measurements, such as positive operator-valued measures (POVMs), as well as imperfections, and facilitates probabilistic quantum information protocols~\cite{Nielsen2010, Peres1990, Jacobs2014, Barnett1998}.

Any linear transformation, including non-unitary ones, can be realized by embedding it into a larger unitary on an extended Hilbert space with ancilla modes representing the environment.  Early frameworks by Knöll and co-workers~\cite{Knll1999, Scheel2000} and by He and colleagues~\cite{He2007} demonstrated this approach but could not address transformations involving both loss and gain. Tischler and co-workers~\cite{Tischler2018} later introduced a quasiunitary dilation method that embeds general linear transformations into physically realizable optical networks using ancilla modes and, when necessary, parametric amplification.

A paradigmatic instance of a non‑unitary optical device is a lossy beam splitter, a 2×2 interferometer with an intrinsic internal absorption coefficient. Such beam splitters exhibit highly tunable, apparently non-linear quantum interference effects~\cite{Vest2017, Barnett1998, Jeffers2000, Uppu2016}.  The lossy beam splitter underpins the non-unitary process of coherent absorption, where a lossy medium partially or fully absorbs coherent light incident from both sides. The absorption depends not only on the medium, but also on the relative phase and amplitudes of the input light. Classically, coherent absorption arises when counterpropagating waves form a standing wave, with the absorption determined by the absorber’s position within the resulting field profile. Complete absorption or coherent perfect absorption (CPA) occurs at antinodes, full transmission at nodes, and partial absorption elsewhere. Coherent perfect absorption, proposed by Chong et al.~\cite{Chong2010} as the time-reversed analogue of a laser, was first experimentally demonstrated using silicon slabs by Wan et al.~\cite{Wan2011}. Since then, classical CPA has been studied for various applications, including light-by-light control~\cite{Zhang2012}, coherent optical switching~\cite{Fang2014}, signal modulation~\cite{Fang2015,Xomalis2018}, dark pulse generation~\cite{Xomalis2018_2}, and all-optical coherent amplification~\cite{Goodarzi2018}.

In the quantum regime, the absorption depends on the input quantum state. Single photon CPA has been experimentally demonstrated in subwavelength films~\cite{Roger2015,Vetlugin2019}, where a photon prepared in a balanced path superposition can be deterministically absorbed or transmitted depending on the input relative path phase. With multiphoton inputs such as two-photon NOON states, CPA induces nonlinear-like behavior even in linear media~\cite{Roger2016,Lyons2019,Jeffers2019,Vetlugin2021}. Theoretical models show that lossy beam splitters can selectively absorb certain superpositions, and suppress or enhance specific Fock components at the output~\cite{Vetlugin2021, Jeffers2019, Barnett1998}. This provides mechanisms for state-selective filtering~\cite{Vetlugin2021, Jeffers2000, Hardal2019, Yanikgonul2019}, coherent manipulation of photon correlations~\cite{Roger2016, Vest2017, Lyons2019, Barnett1998, Huang2014}, and realization of the anti-Hong-Ou-Mandel (Anti-HOM) effect~\cite{Vetlugin2022}, with both local~\cite{Roger2015, Vetlugin2019} and nonlocal~\cite{Roger2016, Jeffers2019} control over photon absorption. 
 
Despite significant progress, previous quantum CPA experiments relied on static, unprogrammable components. Absorber parameters were fixed, allowing only limited operating points, while input phase control required moving parts, limiting stability and tuning speed. Here, we overcome these constraints by emulating quantum coherent absorption in a programmable integrated photonic circuit. Our architecture embeds a tunable lossy beam splitter transformation into an 8-mode universal interferometer circuit based on Clements architecture~\cite{Clements2016}. The circuit is synthesized using the quasi-unitary decomposition method proposed by Tischer et al. in Ref.~\cite{Tischler2018} and allow full tunability of all CPA parameters such as the input state phase $\phi$, and the reflectivity $r$, transmissivity $t$, absorptivity $A$, and the internal phase $\phi_{\mathrm{rt}}$ of the lossy beam splitter. We probe the circuit using single-photon balanced dual-path superposition states and two-photon NOON states. The programmable architecture allows access to all previously reported quantum CPA effects, including deterministic absorption, anti-coalescence, photon bunching, and probabilistic two-photon absorption, within a single device. The output photon-count statistics in both cases exhibit high-fidelity agreement with theory, with Bhattacharyya overlaps (a measure of state similarity, where 1 indicates a perfect match) above 0.93 for all configurations and the circuit allows programmed phase sensitivity distribution among the output Fock states. Our results demonstrate programmable ancilla-assisted photonic circuits as practical tools for quantum state engineering, non-unitary quantum simulations, quantum
state filtering and adaptive, reconfigurable, and multiplexed quantum sensing. 

\section*{Results}

\subsection*{Implementation of Non-Unitary transformation}

In this work, we consider port-symmetric lossy beam splitters with identical coefficients for both ports, and such beam splitters are described by a non-unitary $2\times 2$ scattering matrix $S$~\cite{Barnett1998}:
\begin{equation}
\begin{pmatrix}
\hat{a}_{\text{out}} \\
\hat{b}_{\text{out}}
\end{pmatrix}
=
S
\begin{pmatrix}
\hat{a}_{\text{in}} \\
\hat{b}_{\text{in}}
\end{pmatrix},
\quad
S =
\begin{pmatrix}
t & r \\
r & t
\end{pmatrix},
\label{eq:lossy_bs}
\end{equation}
where $t = |t| e^{i\phi_t}$ and $r = |r| e^{i\phi_r}$ are the complex transmission and reflection amplitudes. Throughout this work, we define the internal phase as $\phi_{rt} = \phi_r - \phi_t$ = $\phi_r$ and $\phi_t = 0$. The light absorption is quantified by the intrinsic absorption coefficient $|A|^2 = 1 - |t|^2 - |r|^2$. See Methods for additional constraint equations governing these parameters.

\begin{figure}[h]
\centering
\includegraphics[width=1\textwidth]{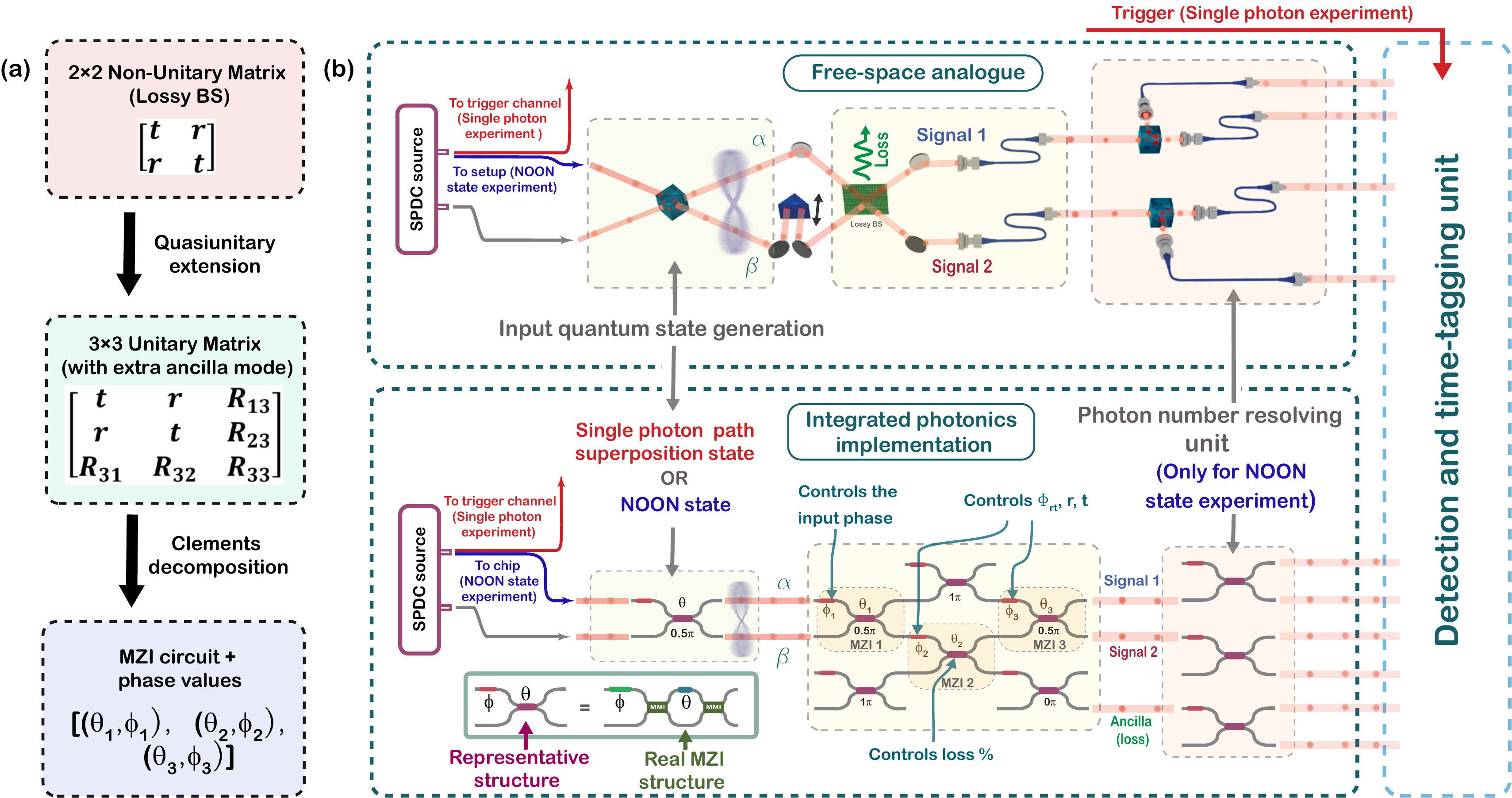}
\caption{%
\textbf{Schematic overview of the coherent perfect absorption experiment.}
\textbf{(a)} Schematic of the theoretical synthesis workflow: a $2\times2$ non-unitary scattering matrix is embedded into a $3\times3$ unitary via quasiunitary extension and decomposed into a Mach-Zehnder interferometer mesh using the Clements scheme. $R_{ij}$ represents the additional complex numbers resulting from the quasiunitary decomposition step.
\textbf{(b)} Conceptual comparison between free-space and integrated photonic implementations of coherent absorption. In both cases, the input is a quantum superposition state over two spatial modes (a single-photon state or a two-photon NOON state), prepared via a bulk 50:50 beam splitter in the free-space setup or an on-chip MZI in the integrated photonic platform. In the free-space case, this state interacts with a static, lossy beam splitter, with absorption governed by the relative phase of the input state. The integrated version emulates this process using a programmable 3-mode interferometer (MZI$_1$–MZI$_3$), enabling tunable control over $|r|$, $|t|$, $\phi_{rt}$, and A of the emulated beam splitter.  All other on-chip MZIs are configured as passive waveguides to route light without affecting the circuit. Detection is performed via heralding (single-photon state case) or photon-number-resolving coincidence counting (NOON state case).
}
\label{fig1}
\end{figure}

Fig.~\ref{fig1}(a) illustrates the workflow used to cast a  $2 \times 2$ non-unitary lossy beam splitter matrix (Eq.~\ref{eq:lossy_bs}) into a larger unitary matrix, enabling its implementation using a linear photonic circuit. We adopt a quasiunitary extension strategy~\cite{Tischler2018}, embedding the lossy transformation into a larger unitary matrix by introducing ancillary vacuum modes (see Methods). We numerically confirmed that, for all beam splitter configurations studied, a single ancilla mode suffices, leading to a compact three-mode implementation of the CPA circuit.

This augmented transformation is then physically synthesized using a rectangular mesh of Mach-Zehnder interferometers (MZIs) via the Clements decomposition scheme~\cite{Clements2016}. Each MZI consists of two fixed 50:50 beam splitters, an internal phase shifter $\theta$ and an external phase shifter $\phi$, together enabling implementation of arbitrary SU(2) transformations. The decomposition determines the phase settings $(\theta_i, \phi_i)$ of each MZI required to implement the full $3 \times 3$ unitary matrix (see Methods).

In this work, we consider circuits emulating two classes of lossy beam splitters characterized by different constraints on their internal parameters. \textbf{Type~1 maintains a fixed internal phase $\bm{\phi_{rt} = \pi}$}, while the magnitude ratio $|r|/|t|$ varies from 0 to 1 as the absorption coefficient $|A|^2$ increases from 0 to 0.5. \textbf{Type~2 imposes symmetry ($\bm{|t| = |r|}$)}, which causes $\phi_{rt}$ to vary from $\pi/2$ to $\pi$ over the same range of $|A|^2$. 

\subsection*{Experimental architecture}

Fig.~\ref{fig1}(b) depicts our experimental architecture alongside a conceptual comparison to free-space CPA using a static lossy beam splitter\cite{Roger2015, Roger2016}. The core transformation is implemented by three reconfigurable MZIs (MZI$_1$–MZI$_3$). Unlike static beam splitters in free-space setup, our circuit allows tunability of $|t|$, $|r|$, $A$, and $\phi_{rt}$.

From a theoretical mapping between the target unitary $S_\text{total}$ and the MZI mesh scattering matrix, we derive compact relations for the MZI phase shifts:
\begin{equation}
\theta_{\mathrm{MZI_2}} = 2 \cos^{-1}\left( \sqrt{2|A|^2} \right),
\label{eq:MZI2theta}
\end{equation}

\begin{equation}
\phi_{\mathrm{MZI_3}} - \phi_{\mathrm{MZI_2}}
= \arg\left( \frac{t + r}{t - r} \right) + \frac{\theta_{\mathrm{MZI_2}}}{2} + \frac{\pi}{2}
= \tan^{-1}\left( \frac{2|t||r|\sin(\phi_{rt})}{|t|^2 - |r|^2} \right) + \frac{\theta_{\mathrm{MZI_2}}}{2} + \frac{\pi}{2}.
\label{eq:phidiff}
\end{equation}

These expressions are consistent with the MZI decomposition schemes and were verified numerically, with detailed derivations in Supplementary Section~S2. We fix $\theta_{\mathrm{MZI_1}} = \theta_{\mathrm{MZI_3}} = 0.5\pi$ (50:50 splitters) and $\phi_{\mathrm{MZI_1}} = \pi$, as these phase values remain the same across the decomposition of all beam splitter configurations. All other on-chip MZIs are set to the passive waveguide modes $\theta = \pi$ (bar state) or $\theta = 0$ (cross state), depending on the desired light routing through the circuit. As evident from Eq.~\ref{eq:MZI2theta}, $\theta_{\mathrm{MZI_2}}$ sets the absorption coefficient $|A|^2$ of the emulated beam splitter, controlling the fraction of light diverted to the ancilla output (mode 3).

We probe this circuit with two classes of quantum input states: single-photon balanced path superposition states and two-photon NOON states. Both states are initialized using a single tunable MZI (depicted on the bottom panel of Fig.~\ref{fig1}(b), configured to perform a balanced $50{:}50$ beam splitter operation.

For the single-photon case, an input photon is split into a superposition across the two signal modes that form the inputs to the circuit:

\begin{equation}
\ket{\Psi_{\mathrm{1ph}}} = \frac{1}{\sqrt{2}} \left( e^{i\phi}\ket{10} - \ket{01} \right),
\label{eq:singlephoton}
\end{equation}

For the 2-photon NOON-state case, two indistinguishable photons injected into the same MZI produce another path-entangled state: 

\begin{equation}
\ket{\Psi_{\mathrm{NOON}}} = \frac{1}{\sqrt{2}} \left( e^{2i\phi} \ket{20} - \ket{02} \right).
\label{eq:NOON}
\end{equation}

In both cases the relative mode phases $\phi, 2\phi$ control the interference with the effective lossy beam splitter. The quantum state phase $\phi$ is applied on top of the fixed $\pi$ offset of $\phi_{\mathrm{MZI_1}}$. The doubled phase dependence ($2\phi$) in case of NOON state enables enhanced phase sensitivity.  

\begin{figure}[h]
\centering
\includegraphics[width=1\textwidth]{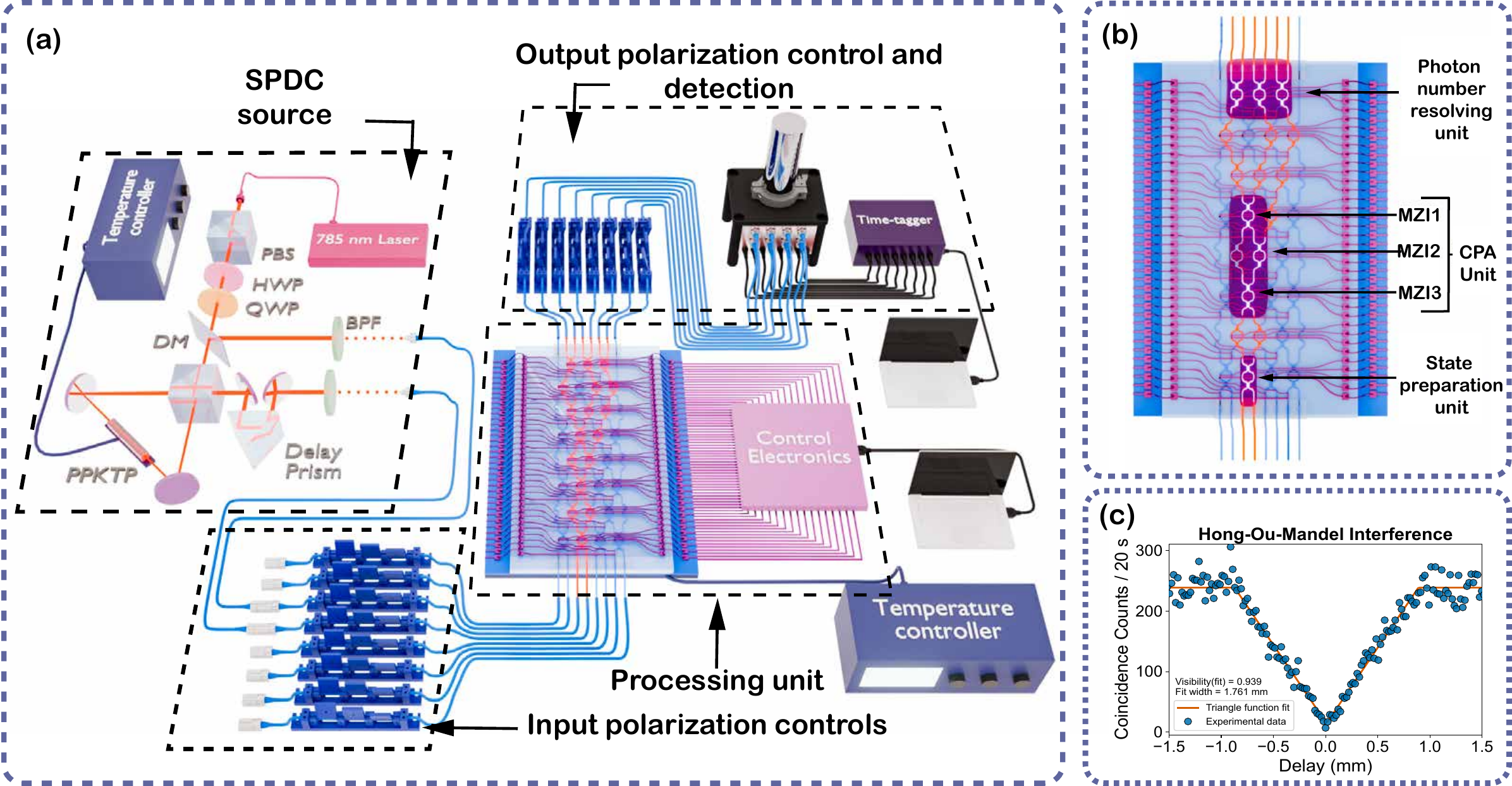}
\caption{\textbf{Experimental setup and photonic circuit layout.}
(a) Schematic of the experimental setup. Photon pairs are generated via type-II spontaneous parametric down-conversion (SPDC) in a PPKTP crystal pumped by a 785\ nm CW laser. Light is coupled in and out of the chip using edge couplers. Fiber-based polarization controllers are used before the input to ensure efficient coupling to the TE mode, and after the chip to rotate the output light to the polarization for which the detectors show maximum efficiency. Output photons are detected using superconducting nanowire single-photon detectors (SNSPDs) (in a 2.6 K cryostat), and coincidence counts are recorded with a Swabian Time Tagger. (b) Functional layout of the $8 \times 8$ programmable photonic chip used in the experiment. The mesh comprises three active subnetworks: (i) a single-MZI unit for preparing single-photon superposition or NOON states, (ii) a central three-mode programmable interferometer implementing the CPA transformation, and (iii) a photon-number-resolving array of MZIs for state analysis. (c) On-chip Hong-Ou-Mandel interference corresponding to NOON state preparation, fitted using a triangular function $y = a - b|x - x_0|$, shows a visibility of $0.939$. This interference dip is obtained by tuning the relative delay between the photon pair paths using a prism mounted on a translational stage on one SPDC output path.
}\label{fig2}
\end{figure}

Fig.~\ref{fig2}(a) illustrates the complete experimental setup used in this work. Correlated photon pairs are generated via type-II spontaneous parametric down-conversion (SPDC) in a periodically poled KTP (PPKTP) crystal, and the down-converted photons, centered around 1570\,nm, are collected into single-mode fibers and routed to the input of the integrated photonic chip.

The chip as a whole comprises an $8 \times 8$ interferometer mesh based on the Clements architecture\cite{Clements2016}, and for this work, we employ three functional subnetworks: the single-MZI state preparation unit, the central 3-mode CPA core, and a photon-number-resolving MZI array for output analysis, as shown in Fig.~\ref{fig2}(b).

All output modes are connected to superconducting nanowire single-photon detectors (SNSPDs), and each detector is assigned to a separate channel of a Swabian Time Tagger, which records and processes detection events with sub-nanosecond timing resolution. Coincidence measurements across all output channels are performed to reconstruct output statistics. In the NOON state experiment, an additional photon-number-resolving step is implemented by routing each of the three output modes from the chip through a fixed 50:50 MZI, effectively splitting each mode into two detection channels(Fig.~\ref{fig1}(b) and Fig.~\ref{fig2}(a)). This enables coincidence detection across all $\binom{6}{2} = 15$ detector combination pairs, from which two-photon output statistics are reconstructed.  For output fock states containing two photons in a single mode (e.g. $\ket{200}$, $\ket{020}$, $\ket{002}$), the measured coincidence counts were multiplied by 2 to account for the 50 \% probability that two photons entering the photon-number-resolving MZI are split into a $\ket{11}$ output, rather than remaining in a $\ket{20}$ or $\ket{02}$ state which yields no coincidence detection. In contrast, for the single-photon experiment, only one photon from the SPDC pair is injected into the chip, while the other serves as a heralding trigger. Coincidence events between the herald and the three chip outputs are recorded, yielding three coincidence channels. This detection strategy provides full access to phase-dependent transmission, absorption, and redistribution effects in both lossy beam splitter configurations.

A 2 ns coincidence window was employed for all measurements presented in this manuscript. See the Methods section and Supplementary Section~S3 for additional information on the experimental setup.

\subsection*{Single-photon state ($\ket{\Psi_{\mathrm{1ph}}}$) experiment}

\begin{figure}[h]
\centering
\includegraphics[width=\textwidth]{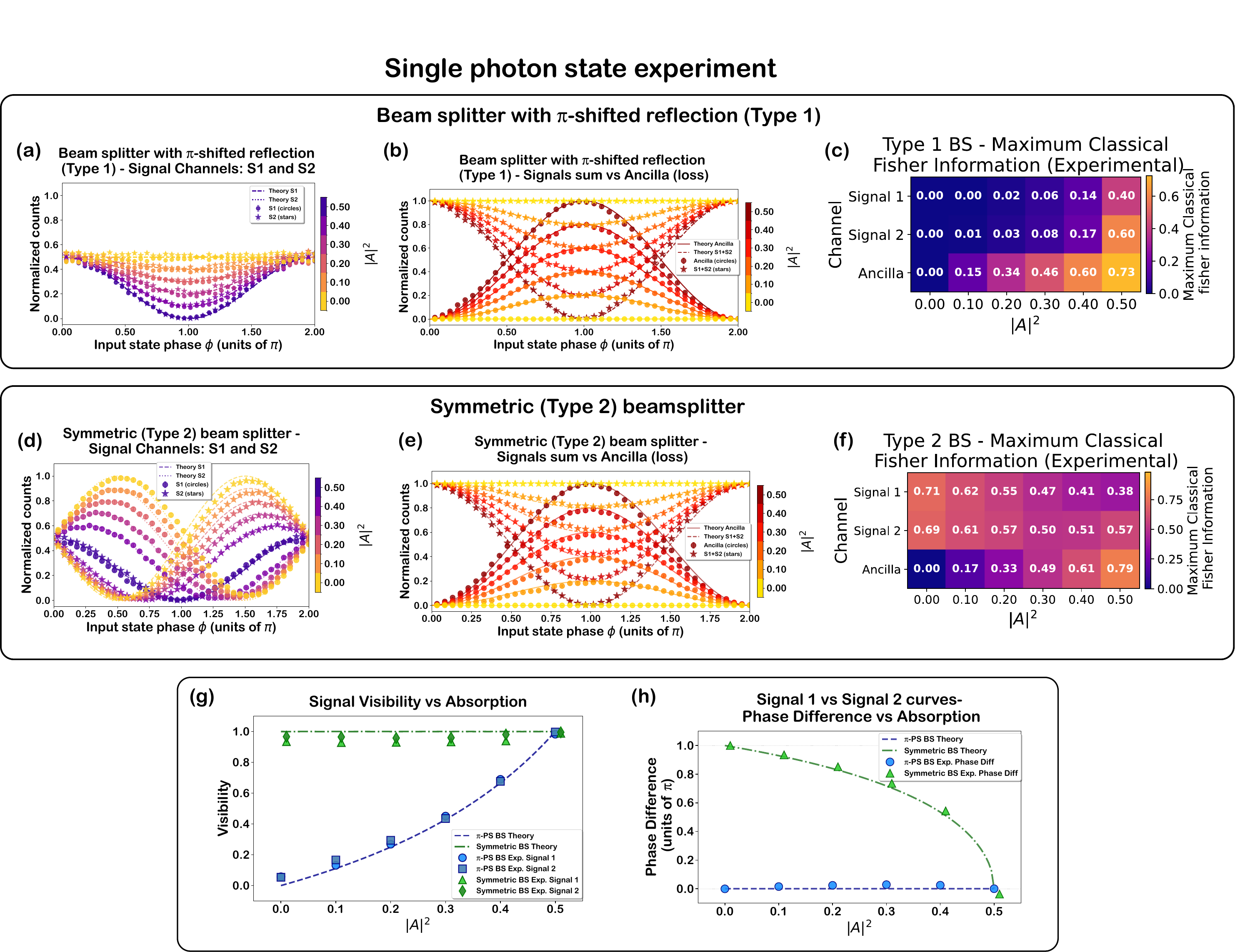}
\caption{\textbf{Single-photon state experiment results}. (a–c) Normalized counts measured at signal (S1, S2) and ancilla (Anc) ports as functions of input phase $\phi$, with the intrinsic absorption coefficient $|A|^2$ represented by the colour scale, for Type~1 ($\phi_{rt} = \pi$) beam splitter. Panels show: (a) individual signal ports; (b) Signals sum (S1+S2) vs. Anc; (c) corresponding Fisher information per output mode for different absorption settings (See Eq.~\ref{eq:fisher_max}). 
(d–f) Same as above for Type~2 symmetric beam splitter ($|t| = |r|$), with (d) signal outputs; (e) Signals sum vs. Anc; (f) Fisher information.  Each data point corresponds to coincidence counts between the circuit outputs and a heralding detector triggered by the twin photon from the SPDC source. All data for both beam splitter configurations are normalized by the total coincidences across all outputs and corrected for variations in detector efficiency and coupling efficiency of the chip’s output waveguides to the corresponding fibers. Error bars represent one standard deviation, calculated as $\sqrt{N}$ from Poisson counting statistics, with proper propagation through the normalization procedure.
(g) Phase-dependent visibility of signal outputs vs. absorption; (h) relative phase shift between S1 and S2 intensities. The lines denote theoretical predictions; symbols are experimental data. The theoretical data are obtained by applying the CPA unitary transformation matrix $S_\text{total}$ to the input state vector and computing output intensities.}
\label{fig3}
\end{figure}

Figure~\ref{fig3} presents measurements with the $\ket{\Psi_{\mathrm{1ph}}}$ state defined in Eq.~\ref{eq:singlephoton}, injected into the two types of beam splitter configurations: Type~1 (fixed $\phi_{rt} = \pi$), and Type~2 ($|t| = |r|$). The single-photon character of the input was verified via a heralded second-order correlation measurement (see Supplementary Section~S4.1), yielding \( g^{(2)}(0) = 0.014 \) and a maximum $g^{(2)}(\tau)$ of only 0.027 within the 2 ns coincidence window. Panels (a–c) and (d–f) respectively correspond to these cases, showing the output intensities at the two signal ports (S1, S2) and the ancilla port (Anc) as functions of input phase $\phi$ and $|A|^2$. 

The observed intensity oscillations arise from coherent single-photon quantum interference at MZI\textsubscript{1} and MZI\textsubscript{3}. For $\phi = 0, 2\pi,\ldots$ (antisymmetric state $\ket{1_{-}} = \frac{1}{\sqrt{2}}(\ket{10} - \ket{01})$), the photon exits deterministically through the upper arm of MZI\textsubscript{1} and reaches the signal ports without absorption. For $\phi = \pi, 3\pi,\ldots$ (symmetric state $-\ket{1_{+}} = -\frac{1}{\sqrt{2}}(\ket{10} + \ket{01})$), the photon exits the lower arm, and gets routed to the ancilla via MZI\textsubscript{2}, with a probability determined by the programmed loss (the set $|A|^2$ value). Unabsorbed photons are routed to the bottom input of MZI\textsubscript{3}. Thus, absorption peaks occur at symmetric input phases. Nearly perfect absorption ($\sim$100\%) is observed at $\phi = \pi$ for $|A|^2 = 0.5$ (Fig.~\ref{fig3}(b,e)). The oscillations follow a $2\pi$ periodicity in input phase.

If the input phase $\phi$ is set independently before $\phi_{\mathrm{MZI_1}}$, applying a $\pi$ shift to $\phi_{\mathrm{MZI_1}}$ swaps the absorbed and transmitted states: the antisymmetric state is absorbed, while the symmetric state is transmitted. For general $\phi$, the input state projects onto both symmetric and antisymmetric modes, resulting in phase-dependent routing between signal and ancilla ports. This behavior is evident in panels Fig.~\ref{fig3}(b, e). The experimental data agree well with the theoretical predictions. Other values of $\phi_{\mathrm{MZI_1}}$ program the circuit to absorb a chosen intermediate superposition state and transmit its orthogonal counterpart. This enables programmable quantum state filtering, selectively routing one superposition component to the ancilla without actually losing the photon.

The internal phase $\phi_{rt}$ governs the relative phase difference between the two inputs to MZI\textsubscript{3} (as described by Eq.~\ref{eq:phidiff}), thereby modulating the interference condition at that MZI. This interference determines how the unabsorbed component of the input state is distributed between the two signal ports. Figure~\ref{fig3}(g) shows that signal port interference visibility increases with absorption for Type~1 but remains fixed at $\sim\ 100\%$ for the symmetric Type~2 configuration. The relative phase shift between the S1 and S2 output intensities, plotted in Fig.~\ref{fig3}(h), stays zero for Type~1 due to its fixed $\phi_{rt} = \pi$, while for Type~2 it evolves continuously with $|A|^2$. The observations in both Fig.~\ref{fig3}(g) and (h) are direct consequences of the $\phi_{rt}$ dependent interference at MZI$_3$.

To quantify phase sensitivity, we evaluate the Classical Fisher information (FI) for each detected output Fock state \( |\psi_i\rangle \).  
For a given lossy BS configuration, the maximum FI with state \( |\psi_i\rangle \) across all input phase values is defined as

\begin{equation}
F_{|\psi_i\rangle}^{\text{max}} =
\max_{\phi} \left[\frac{1}{P_{|\psi_i\rangle}(\phi)} 
\left(
\frac{d P_{|\psi_i\rangle}(\phi)}{d\phi}
\right)^2\right],
\label{eq:fisher_max}
\end{equation}

where \( P_{|\psi_i\rangle}(\phi) \) is the probability of detecting the output fock state \( |\psi_i\rangle \)  \cite{Braunstein1994, Barndorff-Nielsen2000}. The classical Fisher information quantifies the sensitivity of the output state probability to the phase of the input quantum state. The total Fisher information at a given input phase $\phi$ and lossy BS configuration is obtained by summing the contributions of all accessible output Fock states:

\begin{equation}
F_{\mathrm{tot}}(\phi) =
\sum_i 
\frac{1}{P_{|\psi_i\rangle}(\phi)} 
\left(
\frac{d P_{|\psi_i\rangle}(\phi)}{d\phi}
\right)^2.
\label{eq:fisher_total}
\end{equation}

The maximum phase sensitivity of the circuit is then determined from the peak value of \( F_{\mathrm{tot}}(\phi) \) over \( \phi \).

Figure~\ref{fig3}(c, f) plots the maximum FI as a function of $|A|^2$. In both configurations, $F_{|\psi_i\rangle}^{\text{max}}$ peaks in the ancilla port at $|A|^2 = 0.5$ (0.73 for Type~1, 0.79 for Type~2). In both configurations, $F_{\mathrm{tot}}(\phi)$ reaches its maximum value of unity (See Supplementary Fig. S6, S7 (c and d)). For Type 1, the FI across all modes diminishes as the $|A|^2$ decreases. In contrast, for Type~2, phase sensitivity shifts from ancilla to signal ports with reduced loss, illustrating the circuit’s capability for tunable phase-encoded signal routing, redistribution of phase sensitivity among the output modes.

\subsection*{NOON state ($\ket{\Psi_{\mathrm{NOON}}}$) experiment}

\begin{figure}[h]
\centering
\includegraphics[width=0.75\textwidth]{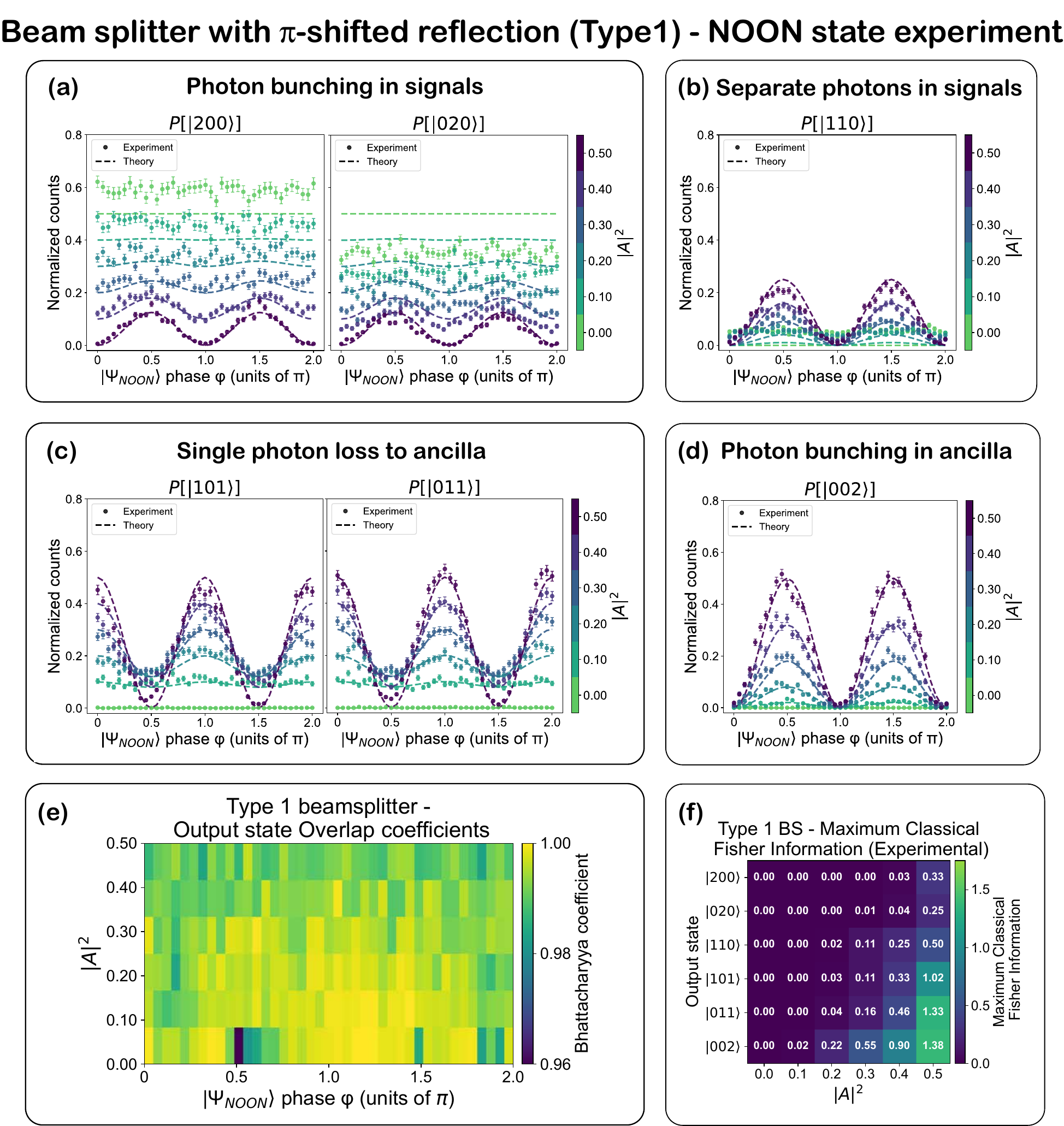}
\caption{\textbf{Beam splitter with $\pi$-shifted reflection (Type1) – NOON state experiment.}  
(a–d) Experimentally measured and normalized counts, and theoretically predicted probabilities for various output Fock states, plotted as a function of the input state phase $\phi$, with $|A|^2$ represented by the colour scale. Probabilities are normalized to the total coincidence counts across all output states and corrected for variations in detector efficiency and coupling efficiency of the chip’s output waveguides to the corresponding fibers. The lines denote theoretical predictions; symbols are experimental data. The theoretical data are obtained by applying the CPA unitary transformation matrix $S_\text{total}$ to the input state vector and computing output intensities. Error bars represent one standard deviation, calculated as $\sqrt{N}$ from Poisson counting statistics, with proper propagation through the normalization procedure.
(e) Bhattacharyya overlap coefficients between experimental and theoretical distributions across $\phi$ and $|A|^2$, evaluated in the Fock basis, demonstrating high-fidelity implementation of the beam splitter transformations. 
(f) Maximum classical Fisher information extracted for each output Fock state at different $|A|^2$ values (See Eq.~\ref{eq:fisher_max}).}
\label{fig4}
\end{figure}

Fig.~\ref{fig2}(c) shows the Hong-Ou-Mandel interference curve measured at the state-preparation MZI used to generate the NOON state. The dip, with a visibility of 93.9\%, confirms high photon indistinguishability, temporal overlap, and state purity. Spectral overlap data for the photon pair (at 1569.5~nm) is provided in Supplementary Section~S4.2.

Fig.~\ref{fig4} and Fig.\ref{fig5} present results for the NOON state experiment with Type~1 and Type~2 beam splitter configurations, respectively. Coincidence counts from all 15 detector pairs are processed to extract photon statistics for the six two-photon Fock basis states: $\ket{200}$, $\ket{020}$, $\ket{110}$, $\ket{101}$, $\ket{011}$, and $\ket{002}$, where the ket is ordered as $\ket{S_1, S_2, \mathrm{Anc}}$.

In the NOON-state case, coherent absorption is governed by programmable two-photon interference at MZI$_1$ and MZI$_3$, and by the programmed single- and two-photon ancilla routing at MZI$_2$. Operating in the quantum regime, these experiments reveal hallmark signatures of genuine multiphoton interference: unlike the classical-like $2\pi$-periodic fringes for single photons, the $\pi$-periodic oscillations here directly reflect the two-photon path entanglement intrinsic to NOON states (Eq.~\ref{eq:NOON}).

\paragraph*{Type 1: Fixed $\phi_{rt} = \pi$.}
In this configuration, the output probabilities of all six Fock basis states exhibit strong phase sensitivity at high $|A|^2$ values. As $|A|^2$ decreases, this phase sensitivity diminishes across all Fock state components, resulting in flatter modulation curves and reduced visibility in the interference patterns. As predicted by theory, at maximal $|A|^2$ and phases $\phi = 0$, $\pi$, and $2\pi$, the system exhibits deterministic single-photon absorption: exactly one photon is routed to the ancilla with near-unity probability, while the remaining photon emerges in either of the two signal modes with equal probability. At the intermediate phases $\phi = 0.5\pi$ and $1.5\pi$, the behavior shifts to probabilistic two-photon absorption, where both photons are absorbed with 50\% probability, and the bunching probability of the absorbed photons approaches 100\%.

The experimental data shown in Fig.~\ref{fig4}(a–d) reflect these trends, highlighting coherent phase-dependent control over quantum interference pathways. Additionally, the probability of photon bunching in the signal modes generally decreases with increasing $|A|^2$, whereas the probability of photon antibunching increases. At intermediate values of $\phi$ and $|A|^2$, the system yields a mixture of all Fock state components, consistent with theoretical predictions for general lossy beam splitters.

In case of the $\ket{200}$ and $\ket{020}$ states, deviations from the ideal curves are observed, particularly at low $|A|^2$ values. This is attributed to directional asymmetry in MZI$_3$ from fabrication imperfections. Although all MZIs are programmed to the desired states, calibration is performed with single-mode inputs, assuming an identical response for the other mode under perfect MMI balance, which may not hold in practice due to fabrication errors. This is similarly evident in the case of the same states in Fig.~\ref{fig5}(a) for the symmetric configuration discussed in the following section.
While slight shifts from the ideal theory lines are present for some output Fock state curves (probably due to cumulative minor imperfections in the overall experimental setup), the overall similarity between the measured and predicted output Fock state probability distributions, quantified by the Bhattacharyya coefficient (Fig.\ref{fig4}(e), Fig.\ref{fig5}(e)) (discussed later), remains high, and the overall trends with varying $|A|^2$ and input phase are consistent with theoretical predictions.

\begin{figure}[h]
\centering
\includegraphics[width=0.75\textwidth]{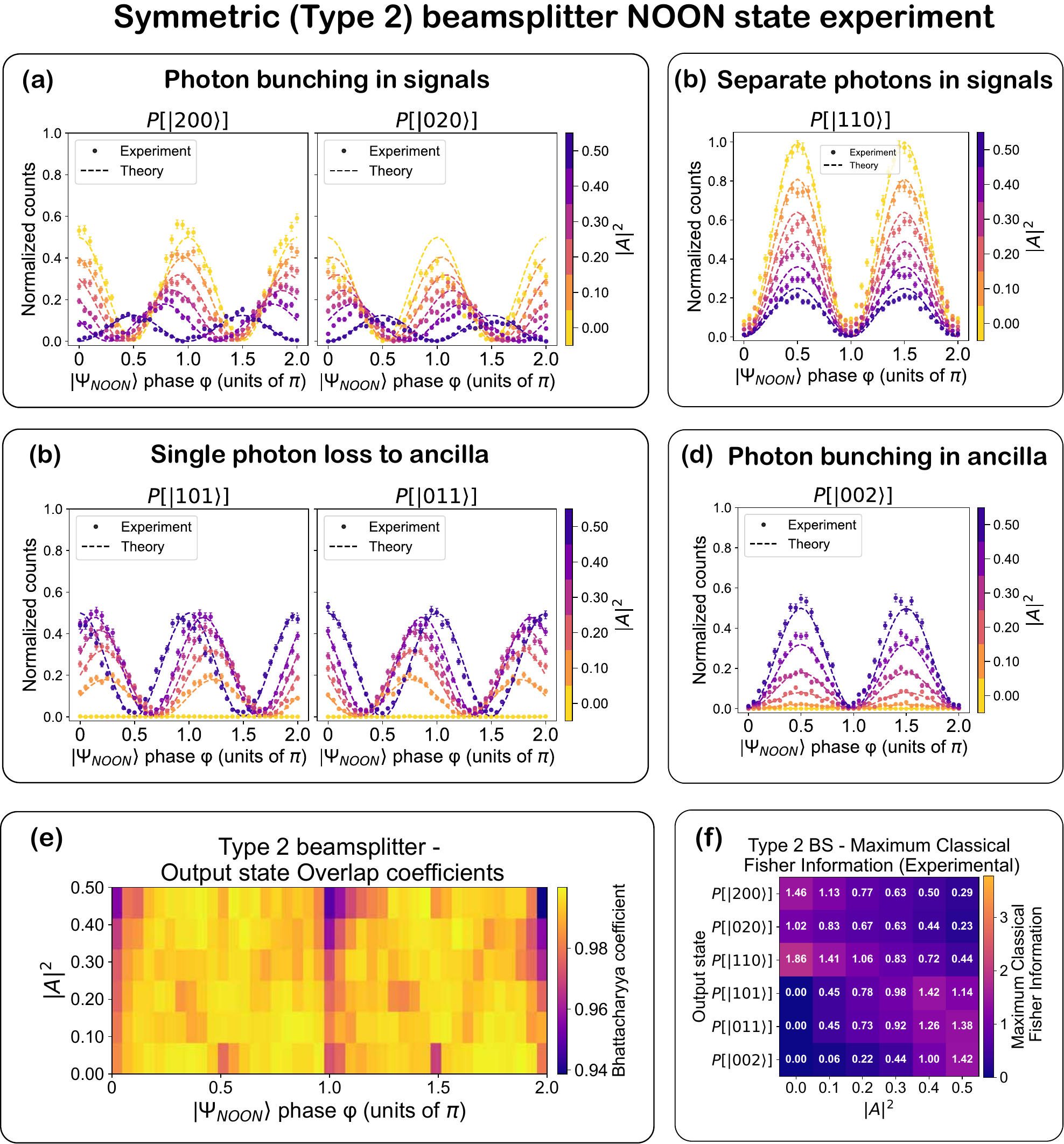}
\caption{\textbf{Symmetric beam splitter (Type 2) – NOON state experiment.} 
(a–d) Experimentally measured and normalized counts, and theoretically predicted probabilities for various output Fock states, plotted as a function of input state phase $\phi$, with $|A|^2$ represented by the colour scale. Probabilities are normalized to the total coincidence counts across all output states and corrected for variations in detector efficiency and coupling efficiency of the chip’s output waveguides to the corresponding fibers.  The lines denote theoretical predictions; symbols are experimental data. The theoretical data are obtained by applying the CPA unitary transformation matrix $S_\text{total}$ to the input state vector and computing output intensities. Error bars represent one standard deviation, calculated as $\sqrt{N}$ from Poisson counting statistics, with proper propagation through the normalization procedure.
(e) Bhattacharyya overlap coefficients between experiment and theory across $\phi$ and $|A|^2$, evaluated in the Fock basis, demonstrating high-accuracy implementation of the beam splitter transformations. 
(f) Maximum classical Fisher information extracted for each output Fock state at different $|A|^2$ values (See Eq.~\ref{eq:fisher_max}).}

\label{fig5}
\end{figure}

\paragraph*{Type 2: Symmetric $|t| = |r|$.}
In the Type~2 configuration, deterministic single-photon absorption occurs at $|A|^2 = 0.5$ and $\phi = 0, \pi, 2\pi$, while probabilistic two-photon absorption with perfect bunching in the ancilla occurs at $\phi = 0.5\pi$ and $1.5\pi$—mirroring the behavior observed in Type~1. However, several key differences emerge. Unlike Type~1, where the phase-dependent modulation of all Fock state probabilities flattens as $|A|^2 \rightarrow 0$, the Type~2 configuration retains significant visibility in several output components even at low absorption. Only the curves of $P(\ket{101})$, $P(\ket{011})$, and $P(\ket{002})$ flatten at $|A|^2 = 0$, while the two photon signal terms ($P(\ket{200})$, $P(\ket{020}), P(\ket{110})$) terms remain phase-sensitive.

At low absorption and $\phi = 0, \pi$, the output exhibits deterministic bunching in the signal ports and at $\phi = 0.5\pi, 1.5\pi$, it switches to deterministic antibunching or anti-coalescence, highlighting the programmable interference nature of the symmetric configuration. A distinguishing feature is the continuous phase shift of the oscillation curves for $P(\ket{200})$, $P(\ket{020})$, $P(\ket{110})$, and $P(\ket{011})$ as the absorption is varied. A full shift of $\pi/2$—half the fringe period—is observed between minimum and maximum absorption values, indicating a dynamic change in the interference landscape. Finally, in contrast to Type~1, both signal bunching and antibunching probabilities decrease with increasing absorption.

\paragraph*{State fidelity and phase sensitivity.} Fig.~\ref{fig4}(e) and  Fig.~\ref{fig5}(e) show the Bhattacharyya overlaps between experimentally measured and theoretically predicted output distributions across the six-dimensional Fock basis. The Bhattacharyya coefficient is defined as
\begin{equation}
B(P, Q) = \sum_{i} \sqrt{p_i q_i},
\label{eq:bhattacharyya}
\end{equation}
where \(P = \{p_i\}\) and \(Q = \{q_i\}\) are the normalized experimental and theoretical probabilities~\cite{Fuchs1999}. A value of \(B = 1\) indicates perfect agreement. Across all absorption and phase settings, we observe overlaps exceeding 0.93, demonstrating high-fidelity implementation of the target transformations. The maximum classical FI $F_{|\psi_i\rangle}^{\text{max}}$ (Fig.~\ref{fig4}(f) and~\ref{fig5}(f)), calculated using Eq.~\ref{eq:fisher_max}, captures the sensitivity of each output mode to input phase changes and the total classical FI, $F_{tot}(\phi)$ calculated using Eq.~\ref{eq:fisher_total}, captures the total phase sensitivity of the circuit to the input state with phase $\phi$. Similar to the single-photon experiment, the Type~1 beam splitters exhibit peaks in classical FI at maximum absorption, particularly in the Fock states with one or two photons in the ancilla ($\ket{002}, \ket{101}, \ket{011}$). As $|A|^2$ decreases, the FI across all Fock states diminishes. In contrast, the Type~2 configuration enables redistribution of phase sensitivity among different Fock components through absorptivity tuning. All Fisher information values reported in this work were extracted from sinusoidal fits to the measured intensity modulations; the fit curves and theoretical predictions are presented in Supplementary Section~S5.

The maximum measured total FI, $F_{tot}(\phi)$ reaches 3.35 for type 1 configuration and 3.4 for type 2 configuration (see supplementary Fig. S8, S9(c and d)), surpassing the shot-noise limit $F_{\mathrm{SQL}} = N = 2$ for two photons and approaching the Heisenberg limit $F_{\mathrm{HL}} = N^2 = 4$, which represents the ultimate quantum precision bound for phase estimation with $N=2$ particles~\cite{Ou1997}.

For the NOON state case also, the circuit can be used to distinguish between orthogonal superposition states. In both configurations, the antisymmetric NOON state $\ket{2_{-}} =  \frac{1}{\sqrt{2}} \left(\ket{20} - \ket{02} \right)$ (corresponding to $\phi = 0, \pi, 2\pi, \ldots$) leads to deterministic single-photon loss into the ancilla. In contrast, the symmetric NOON state $-\ket{2_{+}} = -\frac{1}{\sqrt{2}} \left(\ket{20} + \ket{02} \right)$ (for $\phi = \frac{\pi}{2}, \frac{3\pi}{2}, \ldots$) results in both photons being detected either together in the signal outputs or both directed into the ancilla. As mentioned in an earlier section, the choice of the specific input state that leads to deterministic single-photon absorption can be coherently tuned by adjusting the phase $\phi_{\mathrm{MZI_1}}$.

\section*{Discussion}\label{sec12}

Our experimental results establish a framework for simulating non-unitary quantum transformations using programmable linear optics on an integrated photonic chip. By combining quantum state preparation, programmable circuitry, and photon number resolving detection, we emulate and characterize coherent absorption of quantum light in a fully tunable, ancilla-assisted linear optical architecture.

The two input states — single photon and two photon NOON states — probe distinct interference regimes. The single photon state shows classical coherent absorption like oscillations with $2\pi$ periodicity, while the NOON state exhibits $\pi$ periodic modulation and enhanced phase sensitivity, highlighting the metrological advantage of multiphoton states~\cite{Ou1997}. The peak total classical FI reaches $F_{tot}(\phi) = 1$ for single photon states and $F_{tot}(\phi) = 3.4$ for NOON states, the latter surpassing the shot noise limit ($F = 2$) and approaching the Heisenberg limit ($F = 4$) for two photons.

Comparing the two beam splitter configurations, Type 1 (with fixed $\phi_{rt} = \pi$) imposes a static interference symmetry, while Type 2 (symmetric, $|t| = |r|$) allows dynamic modulation of the internal phase $\phi_{rt}$ through the intrinsic absorptivity. This tunability enables programmable redistribution of phase sensitivity across output modes. In particular, Type 2 facilitates controlled distribution of classical FI among Fock basis components. As a result, each detection outcome carries tailored phase sensitivity, enabling state-selective readout, adaptive quantum sensing, and multiplexed metrology schemes, where phase information is actively steered and shaped within the optical circuit~\cite{Giovannetti2006}.

By transforming a two-mode, two-photon input into a programmable three-mode Fock state distribution, our circuit provides a scalable platform for Fock state engineering and quantum state filtering. Its programmable loss-emulating design is suitable for simulating dissipative channels in quantum thermodynamics and non-Markovian open system dynamics~\cite{Breuer2007}. The three-mode non-unitary block demonstrated here can serve as a reconfigurable subunit in larger photonic quantum processors, enabling more complex functionalities within modular photonic networks. As loss is emulated via an ancilla mode, no photon is destroyed, but rerouted, enabling its independent measurement or reinjection into the main circuit block in a later stage if desired. These capabilities expand the functional repertoire of linear optical systems and open new possibilities for non-unitary quantum photonic protocols.

\section*{Methods}

\subsection*{Lossy beam splitter constraints}

To ensure physical validity, any lossy beam splitter must preserve the canonical bosonic commutation relations, such as $[\hat{a}_{\mathrm{out}},\hat{a}^\dagger_{\mathrm{out}}]=1$ and $[\hat{a}_{\mathrm{out}},\hat{b}^\dagger_{\mathrm{out}}]=0$. For a port-symmetric beam splitter with equal coupling to the loss channel, these constraints lead to an additional phase-dependent relation between $t$, $r$, and $A$: 

\begin{equation}
2|t||r| \cos \phi_{rt} = \pm |A|^2.
\label{eq:phase_condition}
\end{equation}
Together with the energy conservation identity
\begin{equation}
|A|^2 = 1 - |t|^2 - |r|^2,
\label{eq:energy_conservation}
\end{equation}
These equations place a fundamental upper bound on the intrinsic absorption coefficient:
\begin{equation}
|A|^2 \leq 2|t||r| \leq |t|^2 + |r|^2 = 1 - |A|^2 ,
\end{equation}
implying that for a lossy beam splitter, the maximum possible intrinsic absorption coefficient is $|A|^2 = 0.5$.

The procedure used to determine the complex transmission and reflection coefficients t and r from a specified absorption value for both beam splitter types is detailed in Supplementary Section S1.

\subsection*{Quasiunitary decomposition scheme}

To implement a non-unitary transformation using linear optical elements, we employ the quasiunitary extension scheme developed in Ref.~\cite{Tischler2018}. A given non-unitary $2 \times 2$ scattering matrix $S$—representing a lossy beam splitter—is first decomposed via singular value decomposition (SVD):

\begin{equation}
T = U D W,
\end{equation}
where $U$ and $W$ are 2 $\times$ 2 unitary matrices and $D$ is a diagonal matrix of singular values $0 < \sigma_i \leq 1$. Each $\sigma_i < 1$ corresponds to a lossy mode and requires the introduction of an ancilla mode. The total enlarged system has dimension $N = 2 + N_\text{ancilla}$, and the corresponding unitary embedding is constructed as
\begin{equation}
S_{\text{total}} = S_U \cdot S_D \cdot S_W,
\end{equation}
where $S_U$ and $S_W$ are padded versions of $U$ and $W$ acting on the full $N$-mode space, and $S_D$ implements beam splitter-like couplings between lossy signal modes and ancillary vacuum inputs, with attenuation strengths determined by $\sigma_i$.

This construction was implemented numerically using custom Python code that automatically determines the required number of ancilla modes and constructs the enlarged unitary $S_{\text{total}}$ for arbitrary $M \times M$ non-unitary transformations (with $M$ any positive integer, $M=2$ in our case).

For all beam splitter configurations simulated in this work, the quasiunitary decomposition consistently yielded either one or two ancilla modes. However, in every case where two ancilla modes were present, we observed that the second ancilla did not contribute to the final transformation and could be omitted. Thus, the losses could be effectively modeled using a single ancilla mode, consistent with the effective transformation of the beam splitters, and thus making the effective ancilla-embedded transformation a $3\times3$ unitary matrix.

\subsection*{Clements decomposition scheme}

The enlarged $3 \times 3$ unitary $S_{\text{total}}$ is decomposed into a mesh of two-mode unitary operations using the Clements decomposition~\cite{Clements2016}. Each $2 \times 2$ unitary is implemented with a Mach-Zehnder interferometer (MZI), composed of two balanced 50:50 beam splitters and a pair of phase shifters: an internal phase $\theta$ and an external phase $\phi$.

The transfer matrix of a single MZI is expressed as:

\begin{equation}
U_{\text{MZI}}(\theta, \phi) =
e^{i\left(\frac{\theta}{2} + \frac{\pi}{2} \right)}
\begin{pmatrix}
e^{i\phi} \sin\left( \frac{\theta}{2} \right) & \cos\left( \frac{\theta}{2} \right) \\
e^{i\phi} \cos\left( \frac{\theta}{2} \right) & -\sin\left( \frac{\theta}{2} \right)
\end{pmatrix}.
\end{equation}

These MZIs are arranged in a rectangular mesh acting on adjacent modes to synthesize the full unitary transformation. We used a modified version of the open-source \texttt{interferometer} Python package (\href{https://pypi.org/project/interferometer/}{https://pypi.org/project/interferometer/}) to implement the decomposition, adapting its conventions to match our device's MZI transfer matrix. For the CPA transformations used in this work, the Clements mesh contains three MZIs arranged in two layers.

\subsection*{The photonic chip, and thermal stabilization and control}

The photonic circuit is implemented on a silicon-on-insulator platform, featuring silicon waveguides with a height of 250\,nm and width of 450\,nm, fabricated atop a 3\,$\mu$m buried oxide layer. The waveguides are designed to support the fundamental TE mode light at around 1550\,nm. Thermo-optic phase shifters embedded on top of the waveguides provide reconfigurable control over the internal ($\theta$) and external ($\phi$) phase shifts of each MZI. The two beam splitters forming each MZI are balanced 2×2 multimode interference (MMI) couplers.

As illustrated in Fig.~\ref{fig2}, the photonic chip is mounted on a Peltier-cooled stage to suppress thermal drift. The chip is also packaged to ensure stable optical coupling from and to the fiber arrays, in addition to thermal stability. Electrical currents for driving the phase shifters are delivered via wire bonds connected to a custom printed circuit board (PCB) assembly. Each phase shifter is driven using a low-noise current source with PID feedback to ensure precise phase stability.

The SPDC crystal is similarly mounted on a temperature-controlled stage, enabling fine control of its phase-matching condition to generate wavelength-degenerate photon pairs. This ensures high photon indistinguishability—critical for the fidelity of the two-photon NOON state. Before each measurement run, the chip is calibrated using classical light to determine the current–phase mapping of each MZI, ensuring accurate phase programming. For additional details about the experimental setup, refer to Supplementary Section~S3.

\subsection*{Coherent absorption theoretical calculations}

Theoretical phase-dependent intensity curves at various $|A|^2$ values are calculated by applying the corresponding lossy beam splitter scattering matrix to the input state vector via standard matrix multiplication, as defined in Eq.~\ref{eq:matrixmul_single} for the single-photon experiment.
\begin{equation}
\begin{pmatrix}
\sqrt{P_{S_1}} \\
\sqrt{P_{S_2}} \\
\sqrt{P_{\mathrm{Anc}}}
\end{pmatrix}
=
\begin{pmatrix}
t & r & R_{13} \\
r & t & R_{23} \\
R_{31} & R_{32} & R_{33}
\end{pmatrix}
\cdot
\frac{1}{\sqrt{2}}
\begin{pmatrix}
e^{i\phi} \\[2pt]
-1 \\[2pt]
0
\end{pmatrix},
\label{eq:matrixmul_single}
\end{equation}
where the transformation matrix is the enlarged matrix resulting from the quai-unitary extension scheme (see Fig.~\ref{fig1}(a)), $\phi$ is the input single photon quantum state phase, and $P_{S_1}, P_{S_2}$ and $P_{\mathrm{Anc}}$ are the probabilities of corresponding to signal 1, signal 2, and Ancilla, respectively. For the NOON-state predictions, we use the open-source \texttt{qoptcraft} Python package (\href{https://pablovegan.github.io/QOptCraft/\#quantizing-linear-interferomenters}{https://pablovegan.github.io/QOptCraft/\#quantizing-linear-interferomenters}), which maps the $3\times3$ CPA scattering matrix to a $6\times6$ unitary transformation in the two-photon, three-mode Fock basis, following the \textit{generalized bosonic scattering theorem}—also known as the \textit{permanent rule}. This method is implemented in \texttt{qoptcraft} as described in Ref.~\cite{QOptCraft2022}. For clarity, this mapping is illustrated in Eq.~\ref{eq:qoptcraft_mapping}.

\begin{equation}
\underbrace{
\begin{pmatrix}
t & r & R_{13} \\
r & t & R_{23} \\
R_{31} & R_{32} & R_{33}
\end{pmatrix}
}_{\substack{\text{Single-photon, 3 mode}\\\text{$3\times 3$ scattering matrix}}}
\;\xrightarrow[\text{Fock basis conversion}]{\texttt{QOptCraft}}\;
\underbrace{
\begin{pmatrix}
F_{11} & F_{12} & \cdots & F_{16} \\
F_{21} & F_{22} & \cdots & F_{26} \\
\vdots & \vdots & \ddots & \vdots \\
F_{61} & F_{62} & \cdots & F_{66}
\end{pmatrix}
}_{\substack{\text{Two-photon, 3 mode fock basis}\\\text{$6\times 6$ unitary transformation matrix}}}.
\label{eq:qoptcraft_mapping}
\end{equation}

Following the above mapping, the predicted output probability amplitudes for the NOON-state experiment are obtained by multiplying the $6\times6$ unitary with the input state vector in the two-photon, three-mode basis:
\begin{equation}
\begin{pmatrix}
\sqrt{P_{|2,0,0\rangle}} \\[2pt]
\sqrt{P_{|0,2,0\rangle}} \\[2pt]
\sqrt{P_{|1,1,0\rangle}} \\[2pt]
\sqrt{P_{|1,0,1\rangle}} \\[2pt]
\sqrt{P_{|0,1,1\rangle}} \\[2pt]
\sqrt{P_{|0,0,2\rangle}}
\end{pmatrix}
=
\begin{pmatrix}
F_{11} & F_{12} & \cdots & F_{16} \\
F_{21} & F_{22} & \cdots & F_{26} \\
\vdots & \vdots & \ddots & \vdots \\
F_{61} & F_{62} & \cdots & F_{66}
\end{pmatrix}
\cdot
\frac{1}{\sqrt{2}}
\begin{pmatrix}
e^{i2\phi} \\[2pt]
0 \\[2pt]
0 \\[2pt]
-1 \\[2pt]
0 \\[2pt]
0
\end{pmatrix}
\quad
\begin{matrix}
\leftarrow |2,0,0\rangle \\[2pt]
\leftarrow |1,1,0\rangle \\[2pt]
\leftarrow |1,0,1\rangle \\[2pt]
\leftarrow |0,2,0\rangle \\[2pt]
\leftarrow |0,1,1\rangle \\[2pt]
\leftarrow |0,0,2\rangle
\end{matrix},
\label{eq:matrixmul_noon}
\end{equation}

where each $P_{|\cdot\rangle}$ corresponds to the probability of detecting photons in the mode occupation indicated by the ket label $(\text{S1},\text{S2},\text{Anc})$.

\backmatter

\bmhead{Supplementary information}

Supplementary information is available for this paper, including theoretical derivations, additional details on the experimental setup, and additional experimental data.

\bmhead{Acknowledgements}

The authors thank Prof. Ivan M. Khaymovich for helpful discussion. G.K and A.W.E acknowledges the support from Knut and Alice Wallenberg (KAW) Foundation through the Wallenberg Centre for Quantum Technology (WACQT). J.G. acknowledges support from Swedish Research Council (Ref: 2023-06671 and 2023-05288), Vinnova project (Ref: 2024-00466) and the Göran Gustafsson Foundation. A.W.E acknowledges support from Swedish Research Council (VR) Starting Grant (Ref: 2016-03905), and Vinnova quantum kick-start project 2021. V.Z. acknowledges support from the KAW.

\section*{Declarations}
The authors declare no competing financial interests.









\bibliography{sn-bibliography}


\begin{thebibliography}{40}
\ifx \bisbn   \undefined \def \bisbn  #1{ISBN #1}\fi
\ifx \binits  \undefined \def \binits#1{#1}\fi
\ifx \bauthor  \undefined \def \bauthor#1{#1}\fi
\ifx \batitle  \undefined \def \batitle#1{#1}\fi
\ifx \bjtitle  \undefined \def \bjtitle#1{#1}\fi
\ifx \bvolume  \undefined \def \bvolume#1{\textbf{#1}}\fi
\ifx \byear  \undefined \def \byear#1{#1}\fi
\ifx \bissue  \undefined \def \bissue#1{#1}\fi
\ifx \bfpage  \undefined \def \bfpage#1{#1}\fi
\ifx \blpage  \undefined \def \blpage #1{#1}\fi
\ifx \burl  \undefined \def \burl#1{\textsf{#1}}\fi
\ifx \doiurl  \undefined \def \doiurl#1{\url{https://doi.org/#1}}\fi
\ifx \betal  \undefined \def \betal{\textit{et al.}}\fi
\ifx \binstitute  \undefined \def \binstitute#1{#1}\fi
\ifx \binstitutionaled  \undefined \def \binstitutionaled#1{#1}\fi
\ifx \bctitle  \undefined \def \bctitle#1{#1}\fi
\ifx \beditor  \undefined \def \beditor#1{#1}\fi
\ifx \bpublisher  \undefined \def \bpublisher#1{#1}\fi
\ifx \bbtitle  \undefined \def \bbtitle#1{#1}\fi
\ifx \bedition  \undefined \def \bedition#1{#1}\fi
\ifx \bseriesno  \undefined \def \bseriesno#1{#1}\fi
\ifx \blocation  \undefined \def \blocation#1{#1}\fi
\ifx \bsertitle  \undefined \def \bsertitle#1{#1}\fi
\ifx \bsnm \undefined \def \bsnm#1{#1}\fi
\ifx \bsuffix \undefined \def \bsuffix#1{#1}\fi
\ifx \bparticle \undefined \def \bparticle#1{#1}\fi
\ifx \barticle \undefined \def \barticle#1{#1}\fi
\bibcommenthead
\ifx \bconfdate \undefined \def \bconfdate #1{#1}\fi
\ifx \botherref \undefined \def \botherref #1{#1}\fi
\ifx \url \undefined \def \url#1{\textsf{#1}}\fi
\ifx \bchapter \undefined \def \bchapter#1{#1}\fi
\ifx \bbook \undefined \def \bbook#1{#1}\fi
\ifx \bcomment \undefined \def \bcomment#1{#1}\fi
\ifx \oauthor \undefined \def \oauthor#1{#1}\fi
\ifx \citeauthoryear \undefined \def \citeauthoryear#1{#1}\fi
\ifx \endbibitem  \undefined \def \endbibitem {}\fi
\ifx \bconflocation  \undefined \def \bconflocation#1{#1}\fi
\ifx \arxivurl  \undefined \def \arxivurl#1{\textsf{#1}}\fi
\csname PreBibitemsHook\endcsname

\bibitem[\protect\citeauthoryear{Breuer and Petruccione}{2007}]{Breuer2007}
\begin{bbook}
\bauthor{\bsnm{Breuer}, \binits{H.P.}},
\bauthor{\bsnm{Petruccione}, \binits{F.}}:
\bbtitle{The Theory of Open Quantum Systems},
(\byear{2007}).
\doiurl{10.1093/acprof:oso/9780199213900.001.0001}
\end{bbook}
\endbibitem

\bibitem[\protect\citeauthoryear{Reck et~al.}{1994}]{Reck1994}
\begin{barticle}
\bauthor{\bsnm{Reck}, \binits{M.}},
\bauthor{\bsnm{Zeilinger}, \binits{A.}},
\bauthor{\bsnm{Bernstein}, \binits{H.J.}},
\bauthor{\bsnm{Bertani}, \binits{P.}}:
\batitle{Experimental realization of any discrete unitary operator}.
\bjtitle{Physical Review Letters}
\bvolume{73},
\bfpage{58}--\blpage{61}
(\byear{1994})
\doiurl{10.1103/PhysRevLett.73.58}
\end{barticle}
\endbibitem

\bibitem[\protect\citeauthoryear{Clements et~al.}{2016}]{Clements2016}
\begin{barticle}
\bauthor{\bsnm{Clements}, \binits{W.R.}},
\bauthor{\bsnm{Humphreys}, \binits{P.C.}},
\bauthor{\bsnm{Metcalf}, \binits{B.J.}},
\bauthor{\bsnm{Kolthammer}, \binits{W.S.}},
\bauthor{\bsnm{Walsmley}, \binits{I.A.}}:
\batitle{Optimal design for universal multiport interferometers}.
\bjtitle{Optica}
\bvolume{3},
\bfpage{1460}--\blpage{1465}
(\byear{2016})
\doiurl{10.1364/optica.3.001460}
\end{barticle}
\endbibitem

\bibitem[\protect\citeauthoryear{a.~Nielsen and Chuang}{2010}]{Nielsen2010}
\begin{bbook}
\bauthor{\bsnm{Nielsen}, \binits{M.}},
\bauthor{\bsnm{Chuang}, \binits{I.L.}}:
\bbtitle{Quantum Computation and Quantum Information: 10th Anniversary Edition},
(\byear{2010}).
\doiurl{10.1017/CBO9780511976667}
\end{bbook}
\endbibitem

\bibitem[\protect\citeauthoryear{Peres}{1990}]{Peres1990}
\begin{barticle}
\bauthor{\bsnm{Peres}, \binits{A.}}:
\batitle{Neumark's theorem and quantum inseparability}.
\bjtitle{Foundations of Physics}
\bvolume{20},
\bfpage{1441}--\blpage{1453}
(\byear{1990})
\doiurl{10.1007/BF01883517}
\end{barticle}
\endbibitem

\bibitem[\protect\citeauthoryear{Jacobs}{2014}]{Jacobs2014}
\begin{bbook}
\bauthor{\bsnm{Jacobs}, \binits{K.}}:
\bbtitle{Quantum Measurement Theory and Its Applications},
(\byear{2014}).
\doiurl{10.1017/CBO9781139179027}
\end{bbook}
\endbibitem

\bibitem[\protect\citeauthoryear{Barnett et~al.}{1998}]{Barnett1998}
\begin{barticle}
\bauthor{\bsnm{Barnett}, \binits{S.M.}},
\bauthor{\bsnm{Jeffers}, \binits{J.}},
\bauthor{\bsnm{Gatti}, \binits{A.}},
\bauthor{\bsnm{Loudon}, \binits{R.}}:
\batitle{Quantum optics of lossy beam splitters}.
\bjtitle{Phys. Rev. A}
\bvolume{57},
\bfpage{2134}--\blpage{2145}
(\byear{1998})
\doiurl{10.1103/PhysRevA.57.2134}
\end{barticle}
\endbibitem

\bibitem[\protect\citeauthoryear{Knöll et~al.}{1999}]{Knll1999}
\begin{barticle}
\bauthor{\bsnm{Knöll}, \binits{L.}},
\bauthor{\bsnm{Scheel}, \binits{S.}},
\bauthor{\bsnm{Schmidt}, \binits{E.}},
\bauthor{\bsnm{Welsch}, \binits{D.G.}},
\bauthor{\bsnm{Chizhov}, \binits{A.V.}}:
\batitle{Quantum-state transformation by dispersive and absorbing four-port devices}.
\bjtitle{Physical Review A - Atomic, Molecular, and Optical Physics}
\bvolume{59},
\bfpage{4716}--\blpage{4726}
(\byear{1999})
\doiurl{10.1103/PhysRevA.59.4716}
\end{barticle}
\endbibitem

\bibitem[\protect\citeauthoryear{Scheel et~al.}{2000}]{Scheel2000}
\begin{barticle}
\bauthor{\bsnm{Scheel}, \binits{S.}},
\bauthor{\bsnm{Knöll}, \binits{L.}},
\bauthor{\bsnm{Opatrný}, \binits{T.}},
\bauthor{\bsnm{Welsch}, \binits{D.G.}}:
\batitle{Entanglement transformation at absorbing and amplifying four-port devices}.
\bjtitle{Physical Review A - Atomic, Molecular, and Optical Physics}
\bvolume{62},
\bfpage{043803}
(\byear{2000})
\doiurl{10.1103/PhysRevA.62.043803}
\end{barticle}
\endbibitem

\bibitem[\protect\citeauthoryear{He et~al.}{2007}]{He2007}
\begin{barticle}
\bauthor{\bsnm{He}, \binits{B.}},
\bauthor{\bsnm{Bergou}, \binits{J.A.}},
\bauthor{\bsnm{Wang}, \binits{Z.}}:
\batitle{Implementation of quantum operations on single-photon qudits}.
\bjtitle{Physical Review A - Atomic, Molecular, and Optical Physics}
\bvolume{76},
\bfpage{042326}
(\byear{2007})
\doiurl{10.1103/PhysRevA.76.042326}
\end{barticle}
\endbibitem

\bibitem[\protect\citeauthoryear{Tischler et~al.}{2018}]{Tischler2018}
\begin{barticle}
\bauthor{\bsnm{Tischler}, \binits{N.}},
\bauthor{\bsnm{Rockstuhl}, \binits{C.}},
\bauthor{\bsnm{S\l{}owik}, \binits{K.}}:
\batitle{Quantum optical realization of arbitrary linear transformations allowing for loss and gain}.
\bjtitle{Phys. Rev. X}
\bvolume{8},
\bfpage{021017}
(\byear{2018})
\doiurl{10.1103/PhysRevX.8.021017}
\end{barticle}
\endbibitem

\bibitem[\protect\citeauthoryear{Vest et~al.}{2017}]{Vest2017}
\begin{barticle}
\bauthor{\bsnm{Vest}, \binits{B.}},
\bauthor{\bsnm{Dheur}, \binits{M.C.}},
\bauthor{\bsnm{Devaux}},
\bauthor{\bsnm{Baron}, \binits{A.}},
\bauthor{\bsnm{Rousseau}, \binits{E.}},
\bauthor{\bsnm{Hugonin}, \binits{J.P.}},
\bauthor{\bsnm{Greffet}, \binits{J.J.}},
\bauthor{\bsnm{Messin}, \binits{G.}},
\bauthor{\bsnm{Marquier}, \binits{F.}}:
\batitle{Anti-coalescence of bosons on a lossy beam splitter}.
\bjtitle{Science}
\bvolume{356},
\bfpage{1373}--\blpage{1376}
(\byear{2017})
\doiurl{10.1126/science.aam9353}
\end{barticle}
\endbibitem

\bibitem[\protect\citeauthoryear{Jeffers}{2000}]{Jeffers2000}
\begin{barticle}
\bauthor{\bsnm{Jeffers}, \binits{J.}}:
\batitle{Interference and the lossless lossy beam splitter}.
\bjtitle{Journal of Modern Optics}
\bvolume{47},
\bfpage{1819}--\blpage{1824}
(\byear{2000})
\doiurl{10.1080/09500340008232434}
\end{barticle}
\endbibitem

\bibitem[\protect\citeauthoryear{Uppu et~al.}{2016}]{Uppu2016}
\begin{barticle}
\bauthor{\bsnm{Uppu}, \binits{R.}},
\bauthor{\bsnm{Wolterink}, \binits{T.A.W.}},
\bauthor{\bsnm{Tentrup}, \binits{T.B.H.}},
\bauthor{\bsnm{Pinkse}, \binits{P.W.H.}}:
\batitle{Quantum optics of lossy asymmetric beam splitters}.
\bjtitle{Optics Express}
\bvolume{24},
\bfpage{16440}--\blpage{16449}
(\byear{2016})
\doiurl{10.1364/oe.24.016440}
\end{barticle}
\endbibitem

\bibitem[\protect\citeauthoryear{Chong et~al.}{2010}]{Chong2010}
\begin{barticle}
\bauthor{\bsnm{Chong}, \binits{Y.D.}},
\bauthor{\bsnm{Ge}, \binits{L.}},
\bauthor{\bsnm{Cao}, \binits{H.}},
\bauthor{\bsnm{Stone}, \binits{A.D.}}:
\batitle{Coherent perfect absorbers: Time-reversed lasers}.
\bjtitle{Phys. Rev. Lett.}
\bvolume{105},
\bfpage{053901}
(\byear{2010})
\doiurl{10.1103/PhysRevLett.105.053901}
\end{barticle}
\endbibitem

\bibitem[\protect\citeauthoryear{Wan et~al.}{2011}]{Wan2011}
\begin{barticle}
\bauthor{\bsnm{Wan}, \binits{W.}},
\bauthor{\bsnm{Chong}, \binits{Y.}},
\bauthor{\bsnm{Ge}, \binits{L.}},
\bauthor{\bsnm{Noh}, \binits{H.}},
\bauthor{\bsnm{Stone}, \binits{A.D.}},
\bauthor{\bsnm{Cao}, \binits{H.}}:
\batitle{Time-reversed lasing and interferometric control of absorption}.
\bjtitle{Science}
\bvolume{331},
\bfpage{889}--\blpage{892}
(\byear{2011})
\doiurl{10.1126/science.1200735}
\end{barticle}
\endbibitem

\bibitem[\protect\citeauthoryear{Zhang et~al.}{2012}]{Zhang2012}
\begin{barticle}
\bauthor{\bsnm{Zhang}, \binits{J.}},
\bauthor{\bsnm{MacDonald}, \binits{K.F.}},
\bauthor{\bsnm{Zheludev}, \binits{N.I.}}:
\batitle{Controlling light-with-light without nonlinearity}.
\bjtitle{Light Sci. Appl.}
\bvolume{1},
\bfpage{18}
(\byear{2012})
\doiurl{10.1038/lsa.2012.18}
\end{barticle}
\endbibitem

\bibitem[\protect\citeauthoryear{Fang et~al.}{2014}]{Fang2014}
\begin{barticle}
\bauthor{\bsnm{Fang}, \binits{X.}},
\bauthor{\bsnm{Tseng}, \binits{M.L.}},
\bauthor{\bsnm{Ou}, \binits{J.Y.}},
\bauthor{\bsnm{Macdonald}, \binits{K.F.}},
\bauthor{\bsnm{Tsai}, \binits{D.P.}},
\bauthor{\bsnm{Zheludev}, \binits{N.I.}}:
\batitle{Ultrafast all-optical switching via coherent modulation of metamaterial absorption}.
\bjtitle{Applied Physics Letters}
\bvolume{104},
\bfpage{141102}
(\byear{2014})
\doiurl{10.1063/1.4870635}
\end{barticle}
\endbibitem

\bibitem[\protect\citeauthoryear{Fang et~al.}{2015}]{Fang2015}
\begin{barticle}
\bauthor{\bsnm{Fang}, \binits{X.}},
\bauthor{\bsnm{MacDonald}, \binits{K.F.}},
\bauthor{\bsnm{Zheludev}, \binits{N.I.}}:
\batitle{Controlling light with light using coherent metadevices: All-optical transistor, summator and invertor}.
\bjtitle{Light: Science and Applications}
\bvolume{4},
\bfpage{292}
(\byear{2015})
\doiurl{10.1038/lsa.2015.65}
\end{barticle}
\endbibitem

\bibitem[\protect\citeauthoryear{Xomalis et~al.}{2018a}]{Xomalis2018}
\begin{barticle}
\bauthor{\bsnm{Xomalis}, \binits{A.}},
\bauthor{\bsnm{Demirtzioglou}, \binits{I.}},
\bauthor{\bsnm{Plum}, \binits{E.}},
\bauthor{\bsnm{Jung}, \binits{Y.}},
\bauthor{\bsnm{Nalla}, \binits{V.}},
\bauthor{\bsnm{Lacava}, \binits{C.}},
\bauthor{\bsnm{MacDonald}, \binits{K.F.}},
\bauthor{\bsnm{Petropoulos}, \binits{P.}},
\bauthor{\bsnm{Richardson}, \binits{D.J.}},
\bauthor{\bsnm{Zheludev}, \binits{N.I.}}:
\batitle{Fibre-optic metadevice for all-optical signal modulation based on coherent absorption}.
\bjtitle{Nature Communications}
\bvolume{9},
\bfpage{182}
(\byear{2018})
\doiurl{10.1038/s41467-017-02434-y}
\end{barticle}
\endbibitem

\bibitem[\protect\citeauthoryear{Xomalis et~al.}{2018b}]{Xomalis2018_2}
\begin{barticle}
\bauthor{\bsnm{Xomalis}, \binits{A.}},
\bauthor{\bsnm{Demirtzioglou}, \binits{I.}},
\bauthor{\bsnm{Jung}, \binits{Y.}},
\bauthor{\bsnm{Plum}, \binits{E.}},
\bauthor{\bsnm{Lacava}, \binits{C.}},
\bauthor{\bsnm{Petropoulos}, \binits{P.}},
\bauthor{\bsnm{Richardson}, \binits{D.J.}},
\bauthor{\bsnm{Zheludev}, \binits{N.I.}}:
\batitle{Picosecond all-optical switching and dark pulse generation in a fibre-optic network using a plasmonic metamaterial absorber}.
\bjtitle{Applied Physics Letters}
\bvolume{113},
\bfpage{051103}
(\byear{2018})
\doiurl{10.1063/1.5040829}
\end{barticle}
\endbibitem

\bibitem[\protect\citeauthoryear{Goodarzi et~al.}{2018}]{Goodarzi2018}
\begin{barticle}
\bauthor{\bsnm{Goodarzi}, \binits{A.}},
\bauthor{\bsnm{Ghanaatshoar}, \binits{M.}},
\bauthor{\bsnm{Mozafari}, \binits{M.}}:
\batitle{All-optical fiber optic coherent amplifier}.
\bjtitle{Scientific Reports}
\bvolume{8},
\bfpage{15340}
(\byear{2018})
\doiurl{10.1038/s41598-018-33426-7}
\end{barticle}
\endbibitem

\bibitem[\protect\citeauthoryear{Roger et~al.}{2015}]{Roger2015}
\begin{barticle}
\bauthor{\bsnm{Roger}, \binits{T.}},
\bauthor{\bsnm{Vezzoli}, \binits{S.}},
\bauthor{\bsnm{Bolduc}, \binits{E.}},
\bauthor{\bsnm{Valente}, \binits{J.}},
\bauthor{\bsnm{Heitz}, \binits{J.J.F.}},
\bauthor{\bsnm{Jeffers}, \binits{J.}},
\bauthor{\bsnm{Soci}, \binits{C.}},
\bauthor{\bsnm{Leach}, \binits{J.}},
\bauthor{\bsnm{Couteau}, \binits{C.}},
\bauthor{\bsnm{Zheludev}, \binits{N.I.}},
\bauthor{\bsnm{Faccio}, \binits{D.}}:
\batitle{Coherent perfect absorption in deeply subwavelength films in the single-photon regime}.
\bjtitle{Nature Communications}
\bvolume{6},
\bfpage{7031}
(\byear{2015})
\doiurl{10.1038/ncomms8031}
\end{barticle}
\endbibitem

\bibitem[\protect\citeauthoryear{Vetlugin et~al.}{2019}]{Vetlugin2019}
\begin{barticle}
\bauthor{\bsnm{Vetlugin}, \binits{A.N.}},
\bauthor{\bsnm{Guo}, \binits{R.}},
\bauthor{\bsnm{Xomalis}, \binits{A.}},
\bauthor{\bsnm{Yanikgonul}, \binits{S.}},
\bauthor{\bsnm{Adamo}, \binits{G.}},
\bauthor{\bsnm{Soci}, \binits{C.}},
\bauthor{\bsnm{Zheludev}, \binits{N.I.}}:
\batitle{Coherent perfect absorption of single photons in a fiber network}.
\bjtitle{Applied Physics Letters}
\bvolume{115},
\bfpage{191101}
(\byear{2019})
\doiurl{10.1063/1.5118838}
\end{barticle}
\endbibitem

\bibitem[\protect\citeauthoryear{Roger et~al.}{2016}]{Roger2016}
\begin{barticle}
\bauthor{\bsnm{Roger}, \binits{T.}},
\bauthor{\bsnm{Restuccia}, \binits{S.}},
\bauthor{\bsnm{Lyons}, \binits{A.}},
\bauthor{\bsnm{Giovannini}, \binits{D.}},
\bauthor{\bsnm{Romero}, \binits{J.}},
\bauthor{\bsnm{Jeffers}, \binits{J.}},
\bauthor{\bsnm{Padgett}, \binits{M.}},
\bauthor{\bsnm{Faccio}, \binits{D.}}:
\batitle{Coherent absorption of noon states}.
\bjtitle{Physical Review Letters}
\bvolume{117},
\bfpage{023601}
(\byear{2016})
\doiurl{10.1103/PhysRevLett.117.023601}
\end{barticle}
\endbibitem

\bibitem[\protect\citeauthoryear{Lyons et~al.}{2019}]{Lyons2019}
\begin{barticle}
\bauthor{\bsnm{Lyons}, \binits{A.}},
\bauthor{\bsnm{Oren}, \binits{D.}},
\bauthor{\bsnm{Roger}, \binits{T.}},
\bauthor{\bsnm{Savinov}, \binits{V.}},
\bauthor{\bsnm{Valente}, \binits{J.a.}},
\bauthor{\bsnm{Vezzoli}, \binits{S.}},
\bauthor{\bsnm{Zheludev}, \binits{N.I.}},
\bauthor{\bsnm{Segev}, \binits{M.}},
\bauthor{\bsnm{Faccio}, \binits{D.}}:
\batitle{Coherent metamaterial absorption of two-photon states with 40\% efficiency}.
\bjtitle{Phys. Rev. A}
\bvolume{99},
\bfpage{011801}
(\byear{2019})
\doiurl{10.1103/PhysRevA.99.011801}
\end{barticle}
\endbibitem

\bibitem[\protect\citeauthoryear{Jeffers}{2019}]{Jeffers2019}
\begin{barticle}
\bauthor{\bsnm{Jeffers}, \binits{J.}}:
\batitle{Nonlocal coherent perfect absorption}.
\bjtitle{Phys. Rev. Lett.}
\bvolume{123},
\bfpage{143602}
(\byear{2019})
\doiurl{10.1103/PhysRevLett.123.143602}
\end{barticle}
\endbibitem

\bibitem[\protect\citeauthoryear{Vetlugin}{2021}]{Vetlugin2021}
\begin{barticle}
\bauthor{\bsnm{Vetlugin}, \binits{A.N.}}:
\batitle{Coherent perfect absorption of quantum light}.
\bjtitle{Physical Review A}
\bvolume{104},
\bfpage{013716}
(\byear{2021})
\doiurl{10.1103/PhysRevA.104.013716}
\end{barticle}
\endbibitem

\bibitem[\protect\citeauthoryear{Hardal and Wubs}{2019}]{Hardal2019}
\begin{barticle}
\bauthor{\bsnm{Hardal}, \binits{A.U.C.}},
\bauthor{\bsnm{Wubs}, \binits{M.}}:
\batitle{Quantum coherent absorption of squeezed light}.
\bjtitle{Optica}
\bvolume{6}(\bissue{2}),
\bfpage{181}--\blpage{189}
(\byear{2019})
\doiurl{10.1364/OPTICA.6.000181}
\end{barticle}
\endbibitem

\bibitem[\protect\citeauthoryear{Yanikgonul et~al.}{2019}]{Yanikgonul2019}
\begin{bchapter}
\bauthor{\bsnm{Yanikgonul}, \binits{S.}},
\bauthor{\bsnm{Vetlugin}, \binits{A.N.}},
\bauthor{\bsnm{Guo}, \binits{R.}},
\bauthor{\bsnm{Xomalis}, \binits{A.}},
\bauthor{\bsnm{Adamo}, \binits{G.}},
\bauthor{\bsnm{Soci}, \binits{C.}},
\bauthor{\bsnm{Zheludev}, \binits{N.I.}}:
\bctitle{Quantum state filtering of dual-rail photons with fiberized plasmonic metamaterial}.
In: \bbtitle{2019 Conference on Lasers and Electro-Optics, CLEO 2019 - Proceedings}
(\byear{2019}).
\doiurl{10.1364/CLEO_QELS.2019.FTu3D.7}
\end{bchapter}
\endbibitem

\bibitem[\protect\citeauthoryear{Huang and Agarwal}{2014}]{Huang2014}
\begin{barticle}
\bauthor{\bsnm{Huang}, \binits{S.}},
\bauthor{\bsnm{Agarwal}, \binits{G.S.}}:
\batitle{Coherent perfect absorption of path entangled single photons}.
\bjtitle{Optics Express}
\bvolume{22},
\bfpage{20936}--\blpage{20947}
(\byear{2014})
\doiurl{10.1364/oe.22.020936}
\end{barticle}
\endbibitem

\bibitem[\protect\citeauthoryear{Vetlugin et~al.}{2022}]{Vetlugin2022}
\begin{barticle}
\bauthor{\bsnm{Vetlugin}, \binits{A.N.}},
\bauthor{\bsnm{Guo}, \binits{R.}},
\bauthor{\bsnm{Soci}, \binits{C.}},
\bauthor{\bsnm{Zheludev}, \binits{N.I.}}:
\batitle{Anti-hong-ou-mandel interference by coherent perfect absorption of entangled photons}.
\bjtitle{New Journal of Physics}
\bvolume{24},
\bfpage{122001}
(\byear{2022})
\doiurl{10.1088/1367-2630/ac9fe9}
\end{barticle}
\endbibitem

\bibitem[\protect\citeauthoryear{Braunstein and Caves}{1994}]{Braunstein1994}
\begin{barticle}
\bauthor{\bsnm{Braunstein}, \binits{S.L.}},
\bauthor{\bsnm{Caves}, \binits{C.M.}}:
\batitle{Statistical distance and the geometry of quantum states}.
\bjtitle{Physical Review Letters}
\bvolume{72},
\bfpage{3439}--\blpage{3443}
(\byear{1994})
\doiurl{10.1103/PhysRevLett.72.3439}
\end{barticle}
\endbibitem

\bibitem[\protect\citeauthoryear{Barndorff-Nielsen and Gill}{2000}]{Barndorff-Nielsen2000}
\begin{barticle}
\bauthor{\bsnm{Barndorff-Nielsen}, \binits{O.E.}},
\bauthor{\bsnm{Gill}, \binits{R.D.}}:
\batitle{Fisher information in quantum statistics}.
\bjtitle{Journal of Physics A: Mathematical and General}
\bvolume{33}(\bissue{24}),
\bfpage{4481}
(\byear{2000})
\doiurl{10.1088/0305-4470/33/24/306}
\end{barticle}
\endbibitem

\bibitem[\protect\citeauthoryear{Fuchs and van~de Graaf}{1999}]{Fuchs1999}
\begin{barticle}
\bauthor{\bsnm{Fuchs}, \binits{C.A.}},
\bauthor{\bsnm{Graaf}, \binits{J.}}:
\batitle{Cryptographic distinguishability measures for quantum-mechanical states}.
\bjtitle{IEEE Transactions on Information Theory}
\bvolume{45}(\bissue{4}),
\bfpage{1216}--\blpage{1227}
(\byear{1999})
\doiurl{10.1109/18.761271}
\end{barticle}
\endbibitem

\bibitem[\protect\citeauthoryear{Ou}{1997}]{Ou1997}
\begin{barticle}
\bauthor{\bsnm{Ou}, \binits{Z.Y.}}:
\batitle{Fundamental quantum limit in precision phase measurement}.
\bjtitle{Phys. Rev. A}
\bvolume{55},
\bfpage{2598}--\blpage{2609}
(\byear{1997})
\doiurl{10.1103/PhysRevA.55.2598}
\end{barticle}
\endbibitem

\bibitem[\protect\citeauthoryear{Giovannetti et~al.}{2006}]{Giovannetti2006}
\begin{barticle}
\bauthor{\bsnm{Giovannetti}, \binits{V.}},
\bauthor{\bsnm{Lloyd}, \binits{S.}},
\bauthor{\bsnm{Maccone}, \binits{L.}}:
\batitle{Quantum metrology}.
\bjtitle{Phys. Rev. Lett.}
\bvolume{96},
\bfpage{010401}
(\byear{2006})
\doiurl{10.1103/PhysRevLett.96.010401}
\end{barticle}
\endbibitem

\bibitem[\protect\citeauthoryear{Luque et~al.}{2022}]{QOptCraft2022}
\begin{barticle}
\bauthor{\bsnm{Luque}, \binits{A.}},
\bauthor{\bsnm{P\'erez-Salinas}, \binits{A.}},
\bauthor{\bsnm{Mart\'in-L\'opez}, \binits{E.}},
\bauthor{\bsnm{Garc\'ia-Escart\'in}, \binits{J.C.}}:
\batitle{Qoptcraft: A python package for the design and study of linear optical quantum systems}.
\bjtitle{Quantum}
\bvolume{6},
\bfpage{712}
(\byear{2022})
\doiurl{10.22331/q-2022-06-15-712}
\end{barticle}
\endbibitem

\bibitem[\protect\citeauthoryear{Alexiev et~al.}{2021}]{Alexiev2021}
\begin{botherref}
\oauthor{\bsnm{Alexiev}, \binits{C.}},
\oauthor{\bsnm{Mak}, \binits{J.C.C.}},
\oauthor{\bsnm{Sacher}, \binits{W.D.}},
\oauthor{\bsnm{Poon}, \binits{J.K.S.}}:
Calibrating rectangular interferometer meshes with external photodetectors.
OSA Continuum
\textbf{4}
(2021)
\doiurl{10.1364/osac.437918}
\end{botherref}
\endbibitem

\bibitem[\protect\citeauthoryear{Lin et~al.}{2024}]{Lin2024}
\begin{botherref}
\oauthor{\bsnm{Lin}, \binits{S.}},
\oauthor{\bsnm{Zhang}, \binits{Y.}},
\oauthor{\bsnm{Wu}, \binits{Z.}},
\oauthor{\bsnm{Zeng}, \binits{S.}},
\oauthor{\bsnm{Gao}, \binits{Q.}},
\oauthor{\bsnm{Li}, \binits{J.}},
\oauthor{\bsnm{Yu}, \binits{X.}},
\oauthor{\bsnm{Yu}, \binits{S.}}:
Power-efficient programmable integrated multiport photonic interferometer in cmos-compatible silicon nitride.
Photonics Research
\textbf{12}
(2024)
\doiurl{10.1364/prj.507548}
\end{botherref}
\endbibitem

\end{thebibliography}

\newpage

	\baselineskip18pt
	
	\section*{Supplementary Materials: Emulation of Coherent Absorption\\ of Quantum Light in a Programmable Linear\\ Photonic Circuit.}

\noindent
\author
{Govind Krishna,$^{1,\ast}$ Jun Gao,$^{1,\ast}$ Sam O’Brien,$^{1}$ Rohan Yadgirkar,$^{1}$ Venkatesh Deenadayalan,$^{2}$\\ Stefan Preble,$^{2}$ Val Zwiller,$^{1}$ \& Ali W. Elshaari$^{1,\ast}$
	\\
	\\
	\normalsize{$^1$Department of Applied Physics, KTH Royal Institute of Technology, Albanova University Centre, Roslagstullsbacken 21, 106 91 Stockholm, Sweden}\\
    \normalsize{$^2$Microsystems Engineering, Rochester Institute of Technology, Rochester, New York 14623, USA
}\\

    \normalsize{$^\ast$E-mail: govindk@kth.se, junga@kth.se, elshaari@kth.se}\\

}

\baselineskip24pt

\section*{S1. Determination of $t$ and $r$ from a target absorption coefficient $|A|^2$}

We consider a port-symmetric $2\times2$ lossy beamsplitter with scattering matrix
\begin{equation}
S =
\begin{pmatrix}
t & r \\
r & t
\end{pmatrix},
\end{equation}
where $t$ and $r$ are complex transmission and reflection coefficients, respectively.  
The intrinsic absorption coefficient $\alpha := |A|^{2} \in [0,0.5]$ satisfies
\begin{equation}
|t|^{2} + |r|^{2} + \alpha = 1, 
\qquad
2|t||r| \cos\phi_{rt} = \pm \alpha,
\label{eq:abs_constraints}
\end{equation}
with $\phi_{rt} = \arg(r) - \arg(t)$ denoting the internal phase difference.  
We extract $t$ and $r$ from a given $\alpha$ as follows.

\paragraph{Type~1: Fixed $\phi_{rt} = \pi$.}
\begin{enumerate}
\item Set $\cos\phi_{rt} = -1$ in Eq.~\eqref{eq:abs_constraints}.
\item Let $x = |t|^{2}$, so that $|r|^{2} = 1 - \alpha - x$.
\item From Eq.~\eqref{eq:abs_constraints}, obtain
\[
2\sqrt{x(1-\alpha - x)} = \pm\alpha
\quad \Rightarrow \quad
4x(1-\alpha - x) = \alpha^{2}.
\]
\item Solve the quadratic and select the branch
\[
x = \frac{(1-\alpha) + \sqrt{1 - 2\alpha}}{2},
\]
ensuring $|t| \to 1$ as $\alpha \to 0$. If the other branch is selected, we get the solutions satisfying $|r| \to 1$ as $\alpha \to 0$
\item Assign
\[
|t| = \sqrt{x}, \quad |r| = \sqrt{1-\alpha - x}, \quad t = |t|, \quad r = -\,|r| = |r| e^{i\pi}.
\]
\end{enumerate}

\paragraph{Type~2: Equal magnitudes $|t| = |r|$.}
\begin{enumerate}
\item Impose $|t| = |r| = \sqrt{x}$

\item The first constraint gives
\[
2x = 1 - \alpha \quad \Rightarrow \quad x = \frac{1-\alpha}{2}.
\]
\item From the second constraint,  
\[
\cos\phi_{rt} = \frac{\pm \alpha}{2x} = \pm \frac{\alpha}{1-\alpha}.
\]
\item We consider the term with the negative sign and extract $\phi_{rt}$
\[
\phi_{rt} = \arccos\!\left(-\frac{\alpha}{1-\alpha}\right)
, \quad \phi_{rt} \in [0,\pi]\]
\item Assign
\[
t = \sqrt{x}, \quad r = \sqrt{x}\, e^{i\phi_{rt}}.
\]
\end{enumerate}

\paragraph{Remarks.}
For Type~1, the magnitudes of $t$ and $r$ vary with $\alpha$, recovering a fully transparent device ($|t| \to 1$, $|r| \to 0$) as $\alpha \to 0$.  
For Type~2, the magnitudes are fixed at $|t| = |r|$, and it is the relative phase $\phi_{rt}$ that changes with $\alpha$, tending to $\pi/2$ as $\alpha \to 0$.

\section*{S2: Derivation of MZI Phase Conditions}

NB: Throughout this section, we use the notations $\theta_i$ and $\phi_i$ interchangeably with $\theta_{\mathrm{MZI}i}$ and $\phi_{\mathrm{MZI}_i}$ for brevity and to reduce typographical clutter.

\subsubsection*{S2.1 Derivation of $\theta_{\mathrm{MZI_2}}$ from Absorptivity}

MZI\textsubscript{2} in the CPA interferometer sets the effective absorption coefficient $|A|^2$ of the simulated lossy beamsplitter by routing part of the input light into the ancilla mode. As indicated in Fig.1(b), MZI\textsubscript{2} receives input from mode $\hat{a}_{\text{in}}$ and splits it between the signal and ancilla outputs via an SU(2) transformation. Its action on the input can be expressed as:

\begin{equation}
\begin{pmatrix}
\alpha_{\text{S}} \\
\alpha_{\text{anc}}
\end{pmatrix}
=
e^{i\left(\frac{\theta_2}{2} + \frac{\pi}{2} \right)}
\begin{pmatrix}
e^{i\phi_2} \sin\left( \frac{\theta_2}{2} \right) & \cos\left( \frac{\theta_2}{2} \right) \\
e^{i\phi_2} \cos\left( \frac{\theta_2}{2} \right) & -\sin\left( \frac{\theta_2}{2} \right)
\end{pmatrix}
\begin{pmatrix}
\alpha \\
0
\end{pmatrix},
\end{equation}

where $\alpha$ is the complex amplitude of the field entering MZI\textsubscript{2} from MZI\textsubscript{1}, and $\theta_2$ and $\phi_2$ are the internal and external phase shifts of MZI\textsubscript{2}, respectively.

The complex amplitude at the ancilla port is then:

\begin{equation}
\alpha_{\text{anc}} = e^{i\left( \frac{\theta_2}{2} + \frac{\pi}{2} \right)} e^{i\phi_2} \cos\left( \frac{\theta_2}{2} \right) \alpha.
\end{equation}

Taking the intensity at the ancilla output:

\begin{equation}
|\alpha_{\text{anc}}|^2 = |\alpha|^2 \cos^2\left( \frac{\theta_2}{2} \right),
\label{eq:ancilla_intensity}
\end{equation}

We now aim to write the expression for the ancilla intensity entirely in terms of the intrinsic absorption coefficient $|A|^2$, using the beam splitter scattering relations and the constraint equations. The non-unitary transformation of coherent absorption can be expressed in matrix form as:

\begin{align}
&\begin{bmatrix}
t & r \\
r & t
\end{bmatrix}
\times \frac{1}{\sqrt{2}}
\begin{bmatrix}
e^{i\phi} \\
-1
\end{bmatrix}
=
\frac{1}{\sqrt{2}}
\begin{bmatrix}
t e^{i\phi} - r \\
r e^{i\phi} - t
\end{bmatrix}, \nonumber \\[0.5em]
&\text{where $\phi$ is the phase of the input quantum state}\nonumber \\
&\text{Absorbed intensity:} \nonumber \\
&I_{\mathrm{abs}} 
= 1 - \frac{1}{2} \left| t e^{i\phi} - r \right|^2 
       - \frac{1}{2} \left| r e^{i\phi} - t \right|^2 \nonumber \\
&= 1 - \frac{1}{2} \big[ (t e^{i\phi} - r)(t^* e^{-i\phi} - r^*) 
     + (r e^{i\phi} - t)(r^* e^{-i\phi} - t^*) \big] \nonumber \\
&= 1 - \frac{1}{2} \big[ |t|^2 - r t^* e^{-i\phi} - r^* t e^{i\phi} + |r|^2 
     + |r|^2 - t r^* e^{-i\phi} - t^* r e^{i\phi} + |t|^2 \big] \nonumber \\
&= 1 - |t|^2 - |r|^2 
    - \frac{1}{2} \big[ e^{i\phi} + e^{-i\phi} \big] \,\big[ t r^* + r t^* \big] \nonumber \\
&= |A|^2 - \cos\phi \,\big[ t r^* + r t^* \big] \nonumber \\
&= |A|^2 - \cos\phi \,\big[2|t||r|cos\phi_{rt}\big] \nonumber \\
&= |A|^2 \pm |A|^2 \cos\phi.
\end{align}

Thus, at the input state phases $\phi$ corresponding to maximal absorption, the intensity in the ancilla mode, i.e, the absorbed intensity, is given by $2|A|^2$. 
Recall that full light absorption is achieved when $|A|^2 = 0.5$ 
(see the theoretical and experimental curves in Figs. 3(b) and 3(e) of the main text), which is also the fundamental upper bound for the intrinsic absorption coefficient of a port-symmetric lossy beamsplitter (see Methods). 
At these $\phi$ values, the light emerging from MZI$_1$ is directed entirely into the input mode 1 of MZI$_2$. We therefore set Eq.~\ref{eq:ancilla_intensity} equal to $2|A|^2$ at $\alpha =1$ and obtain:

\begin{equation}
2|A|^2 = \cos^2\left( \frac{\theta_2}{2} \right).
\end{equation}

Solving for $\theta_2$, we arrive at the expression:

\begin{equation}
\boxed{
\theta_\mathrm{{MZI_2}}= \theta_{2} = 2 \cos^{-1} \left( \sqrt{2|A|^2} \right),
}
\end{equation}

as used in the main text. This provides a direct mapping between the desired absorptivity A and the internal phase setting of MZI\textsubscript{2}.

\subsubsection*{S2.2 Derivation of the Phase Condition for $\phi_{\mathrm{MZI_3}} - \phi_{\mathrm{MZI_2}}$}

To derive the relative phase condition between the external phase shifters of MZI\textsubscript{2} and MZI\textsubscript{3}, we analytically evaluate the 3-mode unitary transformation implemented by the CPA interferometer circuit. For the ease of calculation, we consider only the transformation of the first two modes (signal1 and signal2). This derivation uses the form of the MZI unitary already introduced in the main manuscript (Eq.13), repeated here for convenience:

\begin{equation}
U_{\text{MZI}}(\theta, \phi) =
e^{i\left(\frac{\theta}{2} + \frac{\pi}{2} \right)}
\begin{pmatrix}
e^{i\phi} \sin\left( \frac{\theta}{2} \right) & \cos\left( \frac{\theta}{2} \right) \\
e^{i\phi} \cos\left( \frac{\theta}{2} \right) & -\sin\left( \frac{\theta}{2} \right)
\end{pmatrix}.
\label{eq:mzi_matrix_supp}
\end{equation}

As mentioned in the main text, throughout all beam splitter circuit configurations, the following phase shift values are constant 
\[
\theta_1 = \theta_3 = \frac{\pi}{2}, \quad \phi_1 = -\pi
\]

Also, we have already shown in the previous section that

\[\theta_2 = 2 \cos^{-1} \left( \sqrt{2|A|^2} \right)
\]

We start by evaluating the Scattering matrices step by step.

\paragraph*{MZI1:}
\begin{equation}
U(\mathrm{MZI}_1) = \frac{C}{\sqrt{2}} \begin{pmatrix} -1 & 1 \\ -1 & -1 \end{pmatrix}.
\end{equation}
where we define: $C = \exp\left(i\frac{3\pi}{4}\right)$.

\paragraph*{MZI2:} Acts only on modes 2 and 3. Here, we define a scattering matrix that describes its action on modes 1 and 2, as these are the modes of interest. The scattering matrix defining this transformation is non-unitary (as some input light is lost to mode 3 at this step and we don't take it into account in this matrix) and is of the form:
\begin{equation}
S(\mathrm{MZI}_2) = \begin{pmatrix}
1 & 0 \\
0 & B
\end{pmatrix},
\end{equation}
where:
\begin{equation}
B = \exp\left[i\left(\frac{\theta_2}{2} + \frac{\pi}{2} \right)\right] e^{i\phi_2}\cdot \sqrt{1-2|A|^2}.
\end{equation}

\paragraph*{MZI3:}
\begin{equation}
U(\mathrm{MZI}_3) = C \cdot \begin{pmatrix} \frac{e^{i\phi_3}}{\sqrt{2}} & \frac{1}{\sqrt{2}} \\
\frac{e^{i\phi_3}}{\sqrt{2}} & -\frac{1}{\sqrt{2}} \end{pmatrix},
\end{equation}

Now we perform the matrix multiplications:

\paragraph*{Step 1: $S_{21} = S(\mathrm{MZI}_2) \cdot U(\mathrm{MZI}_1)$}
\begin{equation}
S_{21} = \begin{pmatrix}
1 & 0 \\
0 & B
\end{pmatrix}
\cdot \frac{C}{\sqrt{2}} \begin{pmatrix} -1 & 1 \\ -1 & -1 \end{pmatrix}
= \frac{C}{\sqrt{2}} \begin{pmatrix} -1 & 1 \\ -B & -B \end{pmatrix}.
\end{equation}

\paragraph*{Step 2: $S_{\text{total}} = U(\mathrm{MZI}_3) \cdot S_{21}$}
\begin{align}
S_{\text{total}} &= C \cdot \begin{pmatrix} \frac{e^{i\phi_3}}{\sqrt{2}} & \frac{1}{\sqrt{2}} \\
\frac{e^{i\phi_3}}{\sqrt{2}} & -\frac{1}{\sqrt{2}} \end{pmatrix} \cdot
\frac{C}{\sqrt{2}}\begin{pmatrix} -1 & 1 \\ -B & -B \end{pmatrix} \\
&= \frac{C^2}{2} \begin{pmatrix}
- e^{i\phi_3} - B & e^{i\phi_3} - B \\
- e^{i\phi_3} + B & e^{i\phi_3} + B
\end{pmatrix}.
\end{align}

\begin{figure}[h]
\centering
\includegraphics[width=1\textwidth]{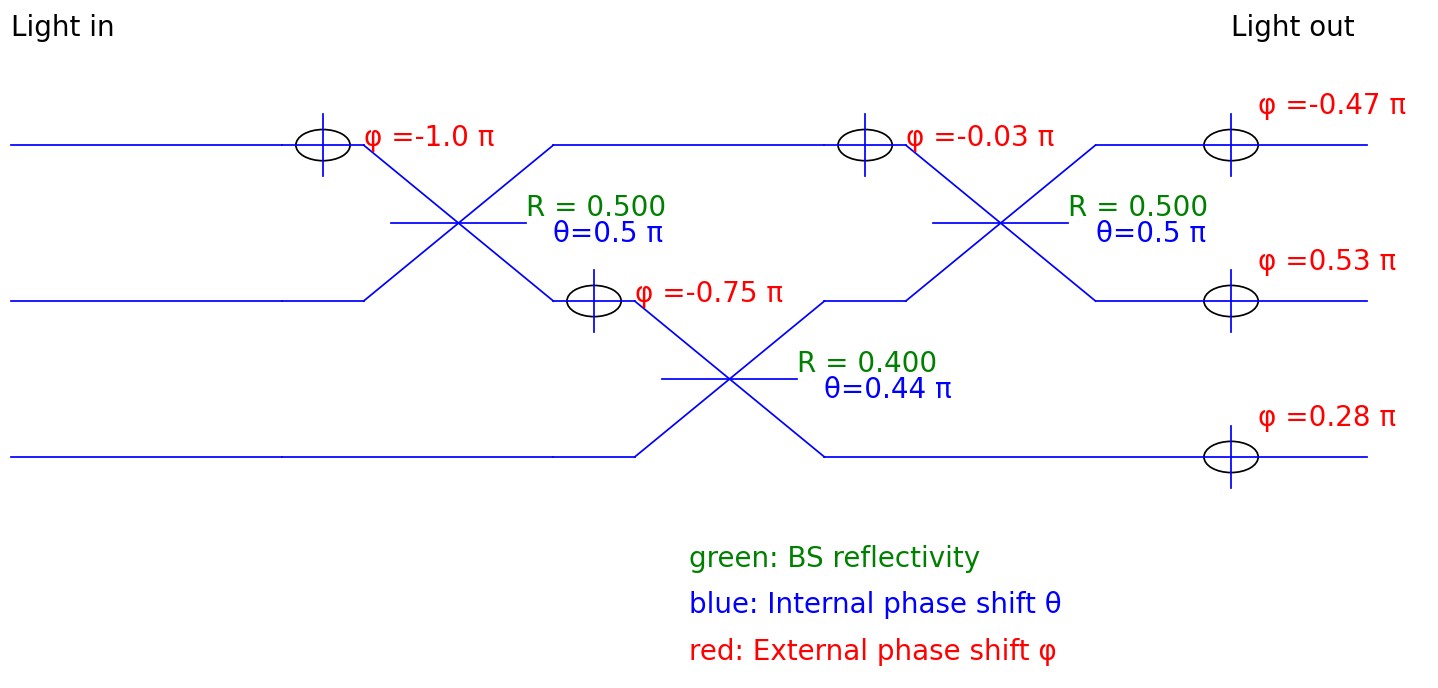}
\caption{Clements decomposition for the CPA circuit corresponding to $|A|^2 = 0.3$ and $\phi_{rt} = \pi$. The decomposition provides internal ($\theta$) and external ($\phi$) phase shift values for each MZI in the mesh, which are used to implement the desired $3\times3$ unitary. The additional output phase shifts visible at the end of each mode (e.g., $\phi = 0.53\pi$ and $\phi = -0.47\pi$) arise from the decomposition and are absorbed into the diagonal matrix $D$ in the theoretical model.}
\label{figS1}
\end{figure}

The Clements decomposition process used to implement this interferometer circuit introduces additional output phases at the end of all output modes of the circuit (see Fig.~\ref{figS1}). These are mode-dependent but consistent across instances of a given decomposition. We denote them by $\delta_1$ and $\delta_2$, which are added to the output of modes 1 and 2, respectively. In our simulations, $\delta_2 - \delta_1 = \pm\pi$ always holds, though the absolute values vary with each decomposition. This was computationally verified by independently performing quasi-unitary extension and Clements decomposition on 10,000 randomly generated port-symmetric lossy beamsplitter matrices.

These phases are modeled as an additional diagonal unitary:
\begin{equation}
D = \mathrm{diag}(e^{i\delta_1}, \; e^{i\delta_2}) = \mathrm{diag}(e^{i\delta_1}, \; e^{i\delta_1 \pm \pi}) = \mathrm{diag}(e^{i\delta_1}, \; -e^{i\delta_1}).
\end{equation}

Thus, the final transformation matrix becomes:
\begin{equation}
S_{\text{eff}} = D \cdot S_{\text{total}} = \begin{pmatrix} e^{i\delta_1} & 0 \\ 0 & -e^{i\delta_1} \end{pmatrix} \cdot S_{\text{total}}.
\end{equation}

\begin{equation}
S_{\text{eff}} = \frac{C^2}{2} \begin{pmatrix}
-e^{i\delta_1}(e^{i\phi_3} + B) & e^{i\delta_1}(e^{i\phi_3} - B) \\
e^{i\delta_1}(e^{i\phi_3} - B) & -e^{i\delta_1}(e^{i\phi_3} + B)
\end{pmatrix}.
\end{equation}

Now, we match this matrix, which is the represents total circuit transformation to the target lossy beamsplitter matrix:
\begin{equation}
S = \begin{pmatrix} t & r \\ r & t \end{pmatrix},
\end{equation}
we identify the first row elements as:
\begin{align}
t &= -\frac{C^2}{2} e^{i\delta_1}(e^{i\phi_3} + B), \\
r &= \frac{C^2}{2} e^{i\delta_1}(e^{i\phi_3} - B).
\end{align}

Adding and subtracting these gives:
\begin{align}
t + r &= -C^2 e^{i\delta_1} B, \\
t - r &= -C e^{i\delta_1} e^{i\phi_3}.
\end{align}

Hence,
\begin{equation}
\frac{t + r}{t - r} = \frac{B}{e^{i\phi_3}}.
\end{equation}

Using the definition of $B$, we substitute:
\begin{equation}
\frac{t + r}{t - r} = \exp[i(\phi_2 - \phi_3)] \cdot \exp\left(i\left( \frac{\theta_{2}}{2} + \frac{\pi}{2} \right)\right)\sqrt{1-2|A|^2}.
\end{equation}

Taking the argument:
\begin{equation}
\boxed{\phi_\mathrm{{MZI_3}} - \phi_\mathrm{{MZI_2}} = \phi_3 - \phi_2 = \arg\left( \frac{t + r}{t - r} \right) + \frac{\theta_\mathrm{{MZI_2}}}{2} + \frac{\pi}{2}}
\end{equation}

We now aim to express the phase difference $\phi_3-\phi_2$ purely in terms of the reflection and transmission amplitudes $|t|$ and $|r|$ and the relative phase shift between them $\phi_{rt}$. $t = |t| e^{i\phi_t}$ and $r = |r| e^{i\phi_r}$. So $\phi_{rt} = \phi_r - \phi_t$ is the relative internal phase of the lossy beamsplitter.

\begin{equation}
\frac{t + r}{t - r} = \frac{|t| e^{i\phi_t} + |r| e^{i\phi_r}}{|t| e^{i\phi_t} - |r| e^{i\phi_r}} = \frac{|t| + |r| e^{i\phi_{rt}}}{|t| - |r| e^{i\phi_{rt}}},
\end{equation}

We compute the argument of the complex ratio using the standard identity for complex division:
\begin{equation}
\arg\left( \frac{a + b e^{i\phi}}{a - b e^{i\phi}} \right)
= \tan^{-1} \left( \frac{2ab \sin\phi}{a^2 - b^2} \right),
\end{equation}
where \( a = |t| \), \( b = |r| \), and \( \phi = \phi_{rt} \).

Applying this identity, we obtain:
\begin{equation}
\arg\left( \frac{t + r}{t - r} \right)
= \tan^{-1} \left( \frac{2 |t| |r| \sin(\phi_{rt})}{|t|^2 - |r|^2} \right).
\end{equation}

Substituting this into the phase condition, we obtain the final compact expression:

\begin{equation}
\boxed{\phi_{\mathrm{MZI_3}} - \phi_{\mathrm{MZI_2}} =
\tan^{-1} \left( \frac{2 |t| |r| \sin(\phi_{rt})}{|t|^2 - |r|^2} \right)
+ \frac{\theta_{\mathrm{MZI_2}}}{2} + \frac{\pi}{2}}
\label{eq:trigform_phi}
\end{equation}

\paragraph*{Note:} All phase shifters on the chip have $2\pi$ periodicity, meaning a phase setting $\phi$ is operationally equivalent to $\phi + 2\pi$. Thus, all phase expressions should be interpreted modulo $2\pi$.

\section*{S3. Additional Experimental Setup Details}

To provide further clarity on device implementation and control electronics, we include the following elaborations:

\subsection*{S3.1 Fabrication and Layout}

\begin{figure}[h]
\centering
\includegraphics[width=\textwidth]{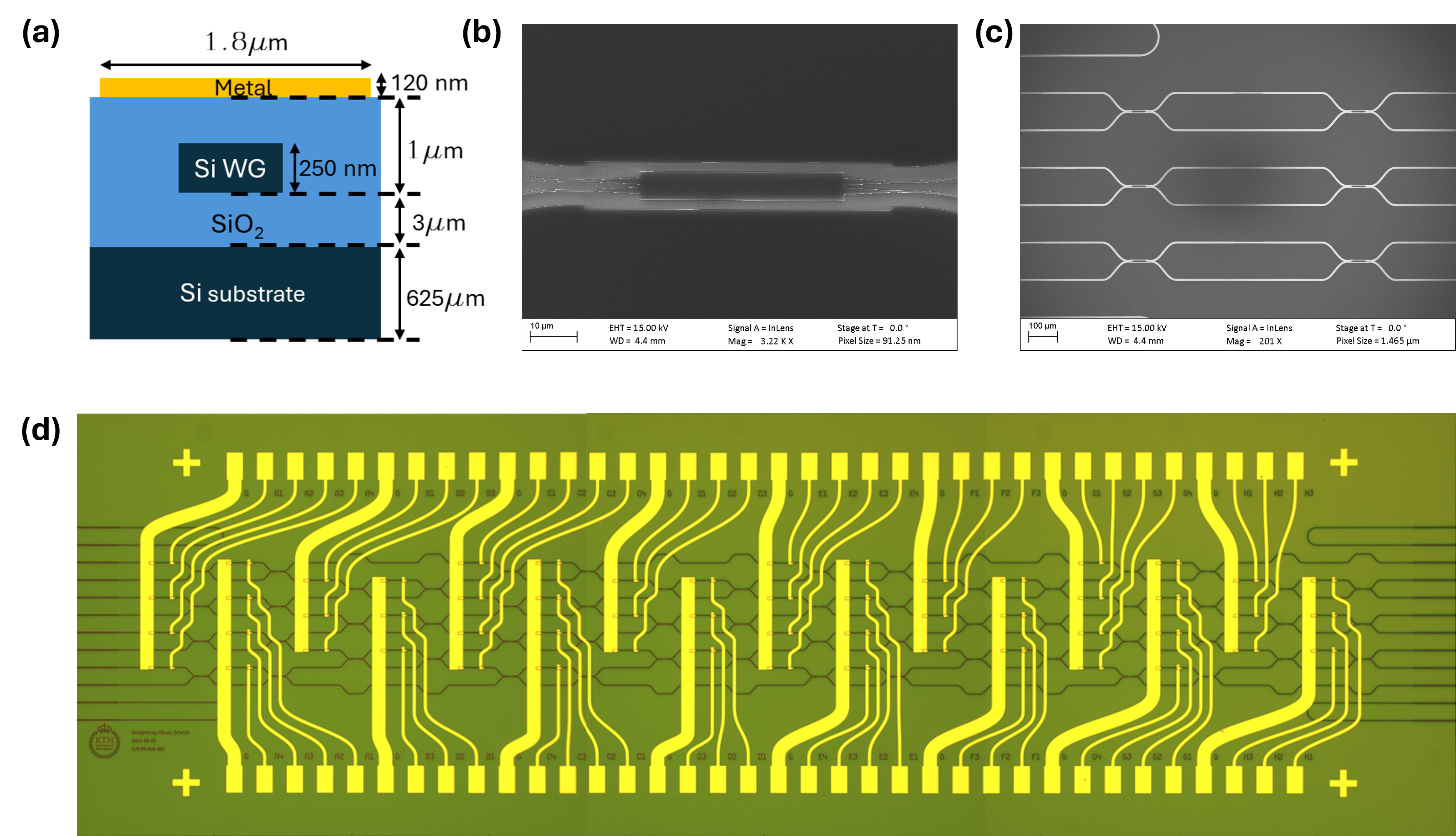}
\caption{
(a) Schematic cross-section of the photonic chip layer stack. The thermo-optic phase shifters are fabricated using a Ti-based metal heater (120\,nm thick, 1.8\,$\mu$m wide), patterned above a 1\,$\mu$m SiO\textsubscript{2} cladding layer to minimize absorption. The underlying waveguide core is a 250 nm silicon layer on 3\,$\mu$m buried oxide atop a 625\,$\mu$m silicon substrate. 
(b) SEM image of a single 2×2 multimode interference (MMI) coupler used as a balanced beamsplitter within each MZI. 
(c) SEM image showing a vertical array of three MZIs forming a single column of the interferometer mesh. 
(d) Stitched optical micrographs of the entire $8 \times 8$ programmable photonic chip, with metal interconnects (yellow) for the thermo-optic phase shifters and edge coupler ports at the chip edge for optical I/O.
}
\label{figS2}
\end{figure}

Fig.~\ref{figS2} provides structural and visual details of the fabricated chip, including the layer stack, SEM images of the core photonic elements, and an optical micrograph of the complete $8 \times 8$ interferometer mesh.

\subsection*{S3.2 Thermo-optic Phase-Shifter Characterization}

The interferometer mesh comprises a total of 56 thermo-optic phase shifters, corresponding to the MZIs in the $8 \times 8$ Clements mesh. Each phase shifter was characterized using continuous-wave laser light at 1550\,nm and automated Python routines. The internal phase shifters were calibrated using the method described in Ref.~\cite{Alexiev2021}, while the external phase shifters followed the calibration scheme outlined in Ref.~\cite{Lin2024}.In both calibration schemes, each phase shifter is embedded within an on-chip interferometric loop, where it modulates the optical output power. The output from a selected port is measured as a function of electrical heating power, and the resulting interference fringe is fitted to extract the fringe visibility and $\pi$-phase switching power.

The I–V characteristics of the heaters were measured and fitted using a third-order polynomial:
\begin{equation}
V(I) = R I + \beta I^3,
\end{equation}
where \( R \) represents the linear resistance and \( \beta \) accounts for nonlinear thermal effects. The nonlinearity arises from the temperature-dependent resistivity of the heater material, which causes the resistance to vary with increasing current.

The optical modulation response was modeled using a cosine function of the applied electrical power:
\begin{equation}
P(V) = A \cdot \cos \left( b \cdot P + c \right) + d,
\end{equation}
where \( A \) is the fringe amplitude, \( b \) is the modulation period (i.e., the inverse of the thermal power required to induce a \(2\pi\) phase shift), \( c \) is the phase offset corresponding to zero input power, and \( d \) represents the vertical offset, typically equal to half the peak-to-peak modulation amplitude. The fringe visibility was calculated as:
\[
\text{Visibility} = \frac{A-d}{d}.
\]

\begin{figure}[h]
\centering
\includegraphics[width=1\textwidth]{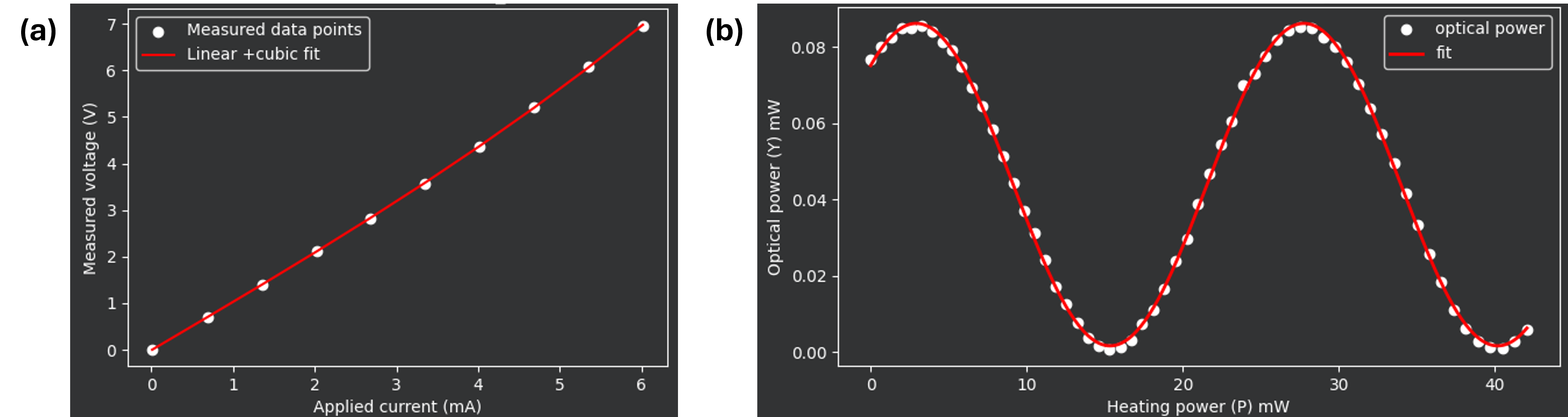}
\caption{Characterization of one thermo-optic phase shifter. \textbf{(a)} Voltage vs.\ current from I-V sweep, showing the linear+cubic fit used to extract the non-linear resistance \textbf{(b)} Optical output power vs.\ applied electrical power, fitted with a cosine-squared function.}
\label{figS3}
\end{figure}

Across all characterized phase shifters, we obtained {\textbf{\boldmath an average experimental fringe visibility of \(0.9984 \pm 0.0003\) and a mean modulation period of \(24.70 \pm 0.03\)~mW\unboldmath}}. An example I-V trace and optical modulation curve from one phase shifter are shown in Fig.~\ref{figS3}, illustrating high fringe contrast and reliable power-to-phase response.

\subsection*{S3.3 Control Electronics}

Thermo-optic phase shifters on the chip are driven using Qontrol Q8-series current driver modules, which provide eight software-defined output channels per module, each capable of delivering up to 24\,mA current and 12\,V voltage. The drivers support both current and voltage control modes with high precision—current output and sensing are accurate to \(\pm 370\,\mathrm{nA}\), and voltage to \(\pm 180\,\mu\mathrm{V}\), with 16-bit output and 18-bit input resolution. All channels are individually programmable via USB through a custom Python interface, enabling automated, scalable control of interferometric meshes and active photonic circuits.

\vspace{1em}
\noindent
\textit{More information:} \href{https://qontrol.co.uk/product/q8iv/}{https://qontrol.co.uk/product/q8iv/}

\subsection*{S3.4 Temperature Stabilization}
Temperature control employed a Thorlabs TED200C benchtop TEC controller (range ±2A / 12W) in conjunction with a TH10K thermistor and TECF1S cooler/heater unit. One controller regulated the SPDC crystal temperature, while a second maintained the photonic chip at a constant \(25^\circ\mathrm{C}\), ensuring stable operation, phase coherence, and wavelength stability.

\section*{S4. SPDC source data}

\subsection*{S4.1 Heralded $g^{(2)}$ measurement of the SPDC source}

\begin{figure}[ht]
    \centering
    \includegraphics[width=0.65\textwidth]{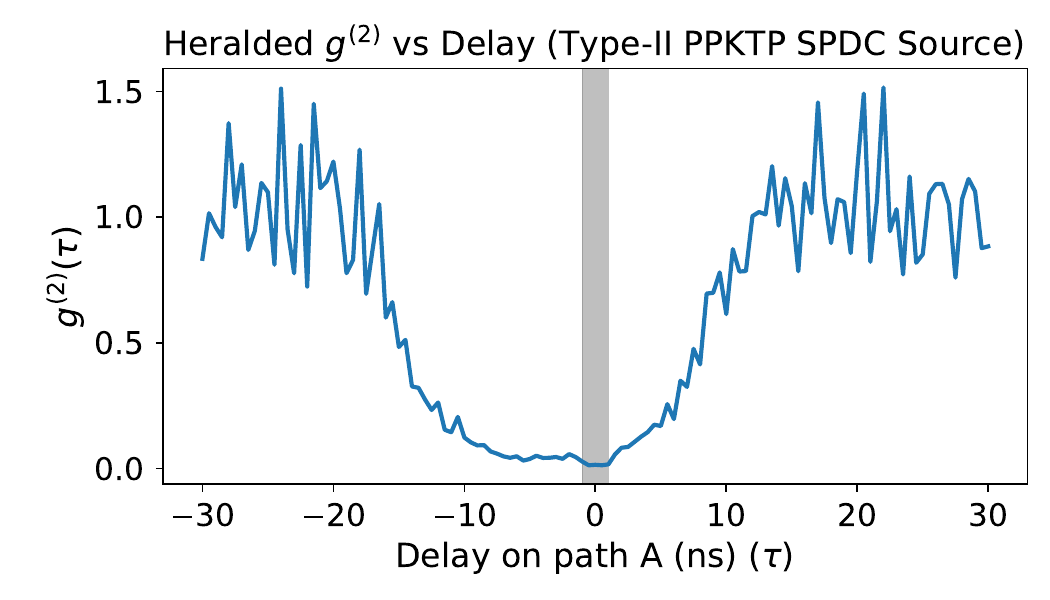}
    \caption{\textbf{Heralded $g^{(2)}$ measurement.} Second-order correlation function $g^{(2)}(\tau)$ of Output 1 from the heralded single-photon source. The shaded grey region around zero delay indicates the 2ns coincidence window used for all time-correlated photon counting measurements throughout this work.}
    \label{fig:g2_heralded}
\end{figure}

To confirm the single-photon nature of our source, we performed a heralded second-order correlation measurement using a standard three-detector Hanbury Brown–Twiss configuration (Fig.~\ref {fig:g2_heralded}). The light from output 1 of the Type-II PPKTP SPDC source was split using a 50:50 fiber beamsplitter and detected at two outputs, labeled A and B, while the idler photon was used as a heralding trigger. The second-order correlation function $g^{(2)}(\tau)$ was computed from the three-fold coincidence counts among all detectors. 

The heralded second-order correlation function is given by
\begin{equation}
g^{(2)}(\tau) = \frac{R_{ABH}(\tau) \, R_H}{R_{AH} \, R_{BH}(\tau)},
\label{eq:g2}
\end{equation}
where $R_{ABH}(\tau)$ is the rate of three-fold coincidences between detectors A, B, and the heralding detector H at time delay $\tau$, $R_{AH}$ and $R_{BH}(\tau)$ are the two-fold coincidence rates between detectors A and H, and B and H respectively, and $R_H$ is the single count rate of the herald. A dip in $g^{(2)}(\tau)$ around $\tau = 0$ is a signature of a single-photon state.

We measured an excellent $g^{(2)}(0) = 0.014$, indicating strong suppression of multi-photon contributions and validating the use of this source for single-photon experiments. The measurement shown in Fig.~\ref{fig:g2_heralded} corresponds to Output 1 of the SPDC source, which was used in all single-photon experiments. The maximum value of $g^{(2)}(\tau)$ within the 2ns coincidence window—highlighted by the shaded grey area in the figure—was found to be 0.0273.

\section*{S4.2 SPDC outputs spectra}

\begin{figure}[ht]
    \centering
    \includegraphics[width=0.75\textwidth]{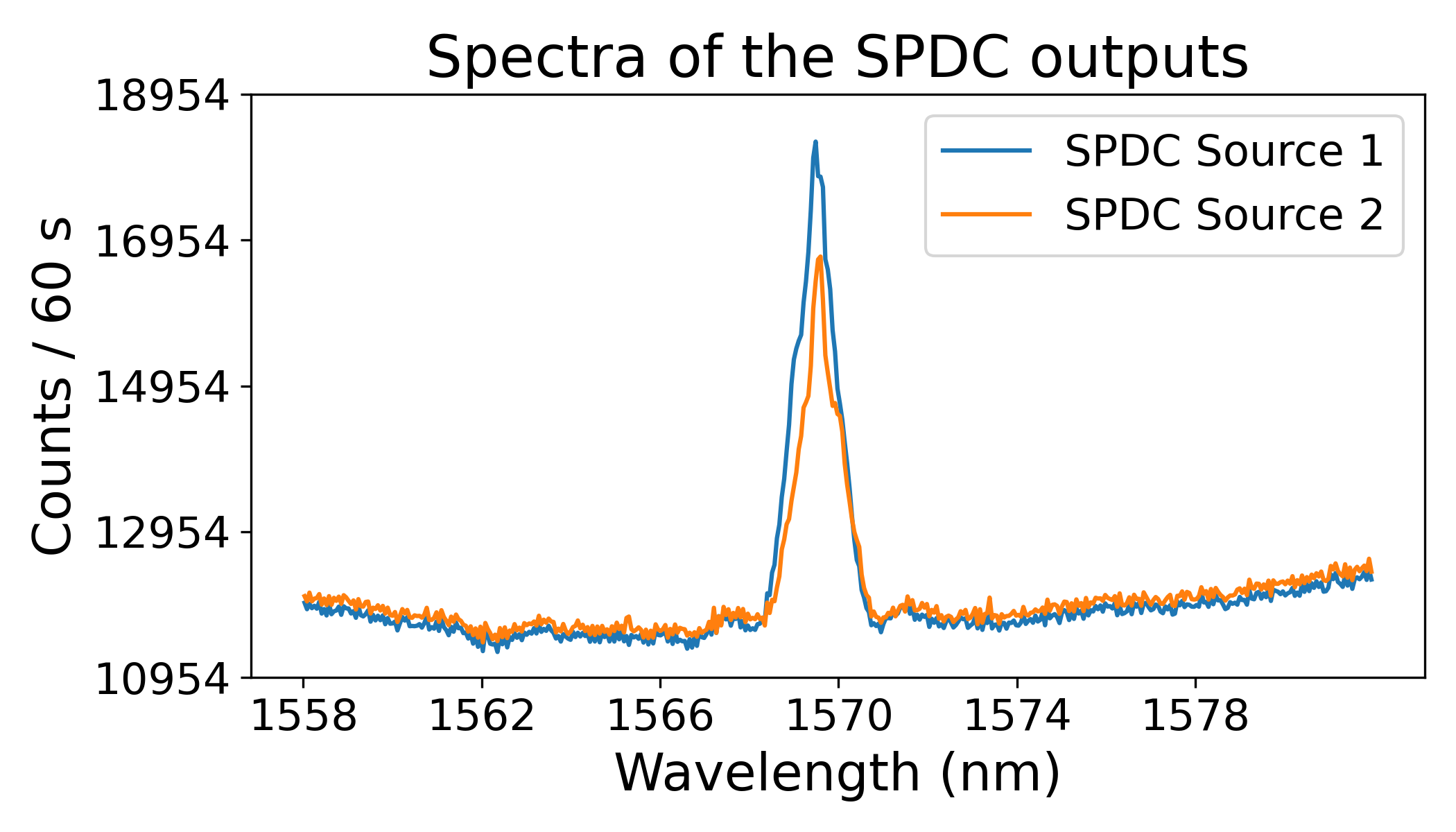}
    \caption{\textbf{Spectral overlap of the SPDC sources.}  
    Measured emission spectra of the two SPDC source outputs used in the experiments, recorded under identical pump and collection conditions. The high degree of spectral overlap ensures the indistinguishability of the photons, which is essential for generating high-fidelity NOON states and the Hong-Ou-Mandel dip. Out 1 peak is centered at 1569.48 nm and Out2 peak at 1569.59 nm. This overlap was achieved by tuning the temperature of the temperature control stage to 28.5$^0$C}
    \label{fig:spdc_spectra}
\end{figure}

\newpage

\section*{S5. Additional details on Maximum Classical Fisher information calculation.}

Additional details of the Fisher information analysis for both single-photon and NOON state experiments are provided in Supplementary Fig.~\textbf{S6--S9}, which show the sinusoidal fits used to extract the phase derivatives and the resulting comparisons between experimental and theoretical maximum classical Fisher information values across all output modes and absorption coefficient settings. The experimentally obtained fisher information heatmaps show good agreement with the theoretically predicted values.

\begin{figure}[h]
\centering
\includegraphics[width=0.8\textwidth]{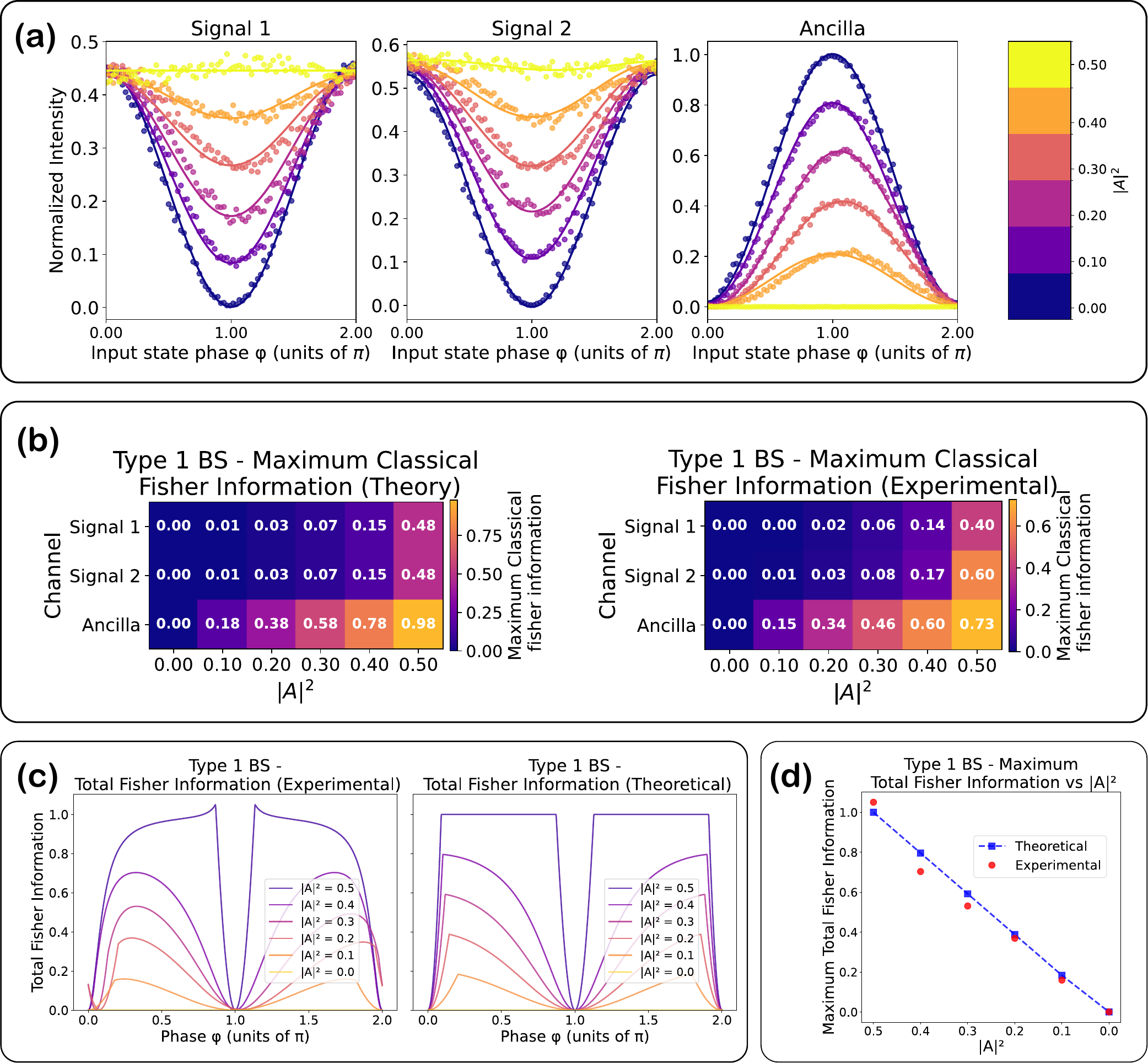}
\caption{\textbf{Fisher information analysis for the single-photon experiment (Type~1 configuration).}
}
\label{figS6}
\end{figure}

\begin{figure}[h]
\centering
\includegraphics[width=0.8\textwidth]{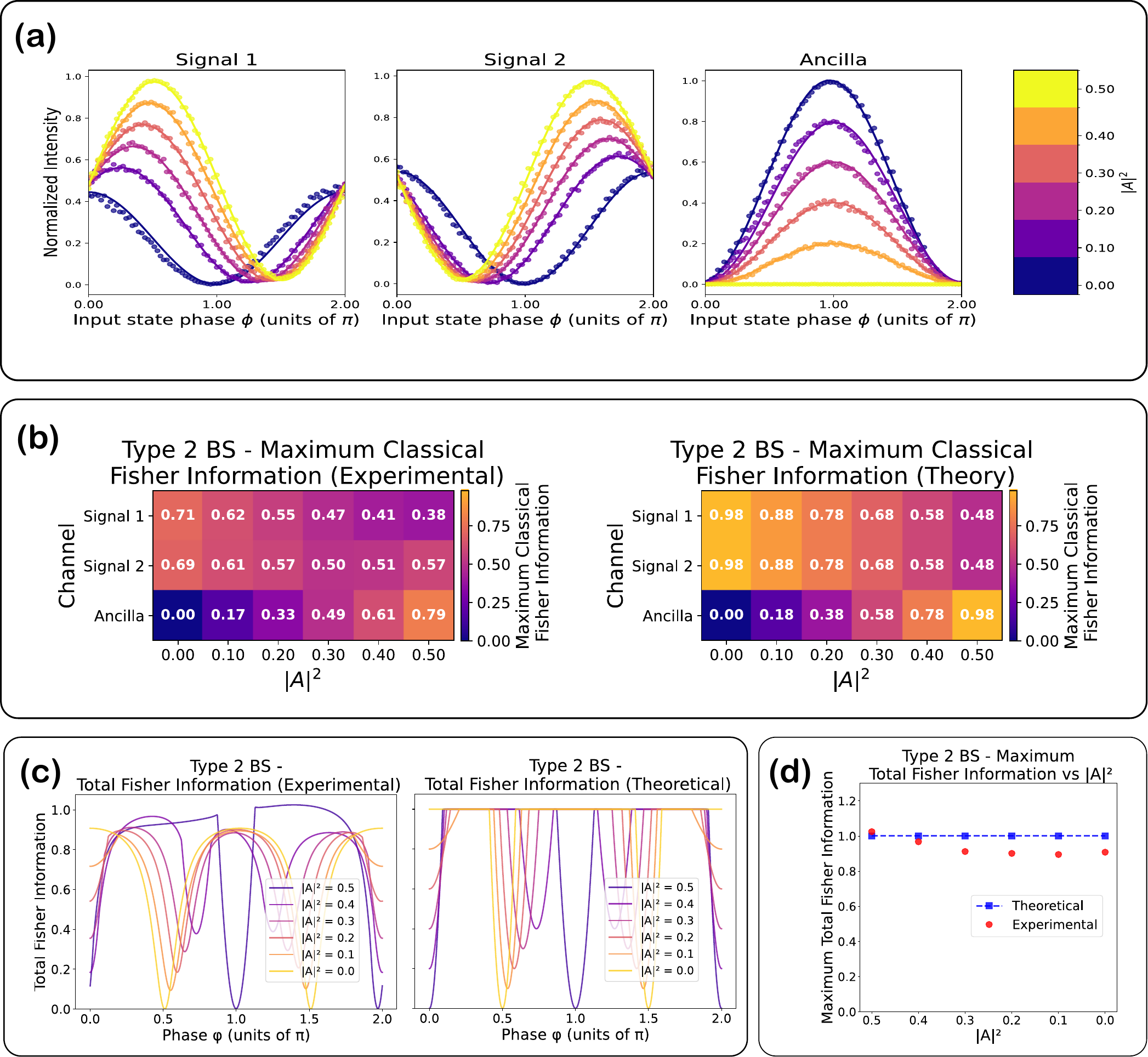}
\caption{\textbf{Fisher information analysis for the single-photon experiment (Type~2 configuration).}
}
\label{figS7}
\end{figure}

\begin{figure}[h]
\centering
\includegraphics[width=0.8\textwidth]{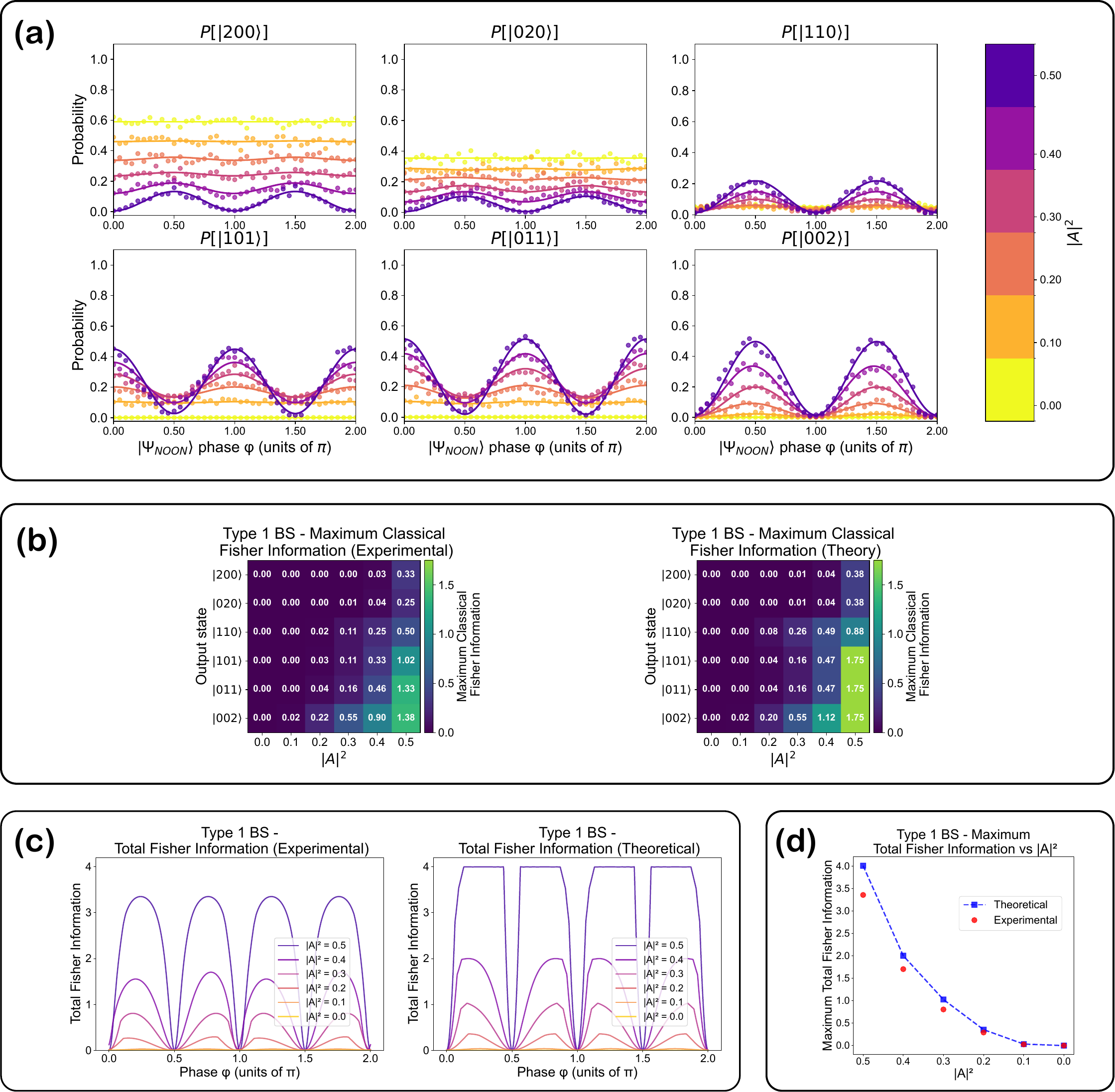}
\caption{\textbf{Fisher information analysis for the NOON state experiment (Type~1 configuration).}
}
\label{figS8}
\end{figure}

\begin{figure}[h]
\centering
\includegraphics[width=0.8\textwidth]{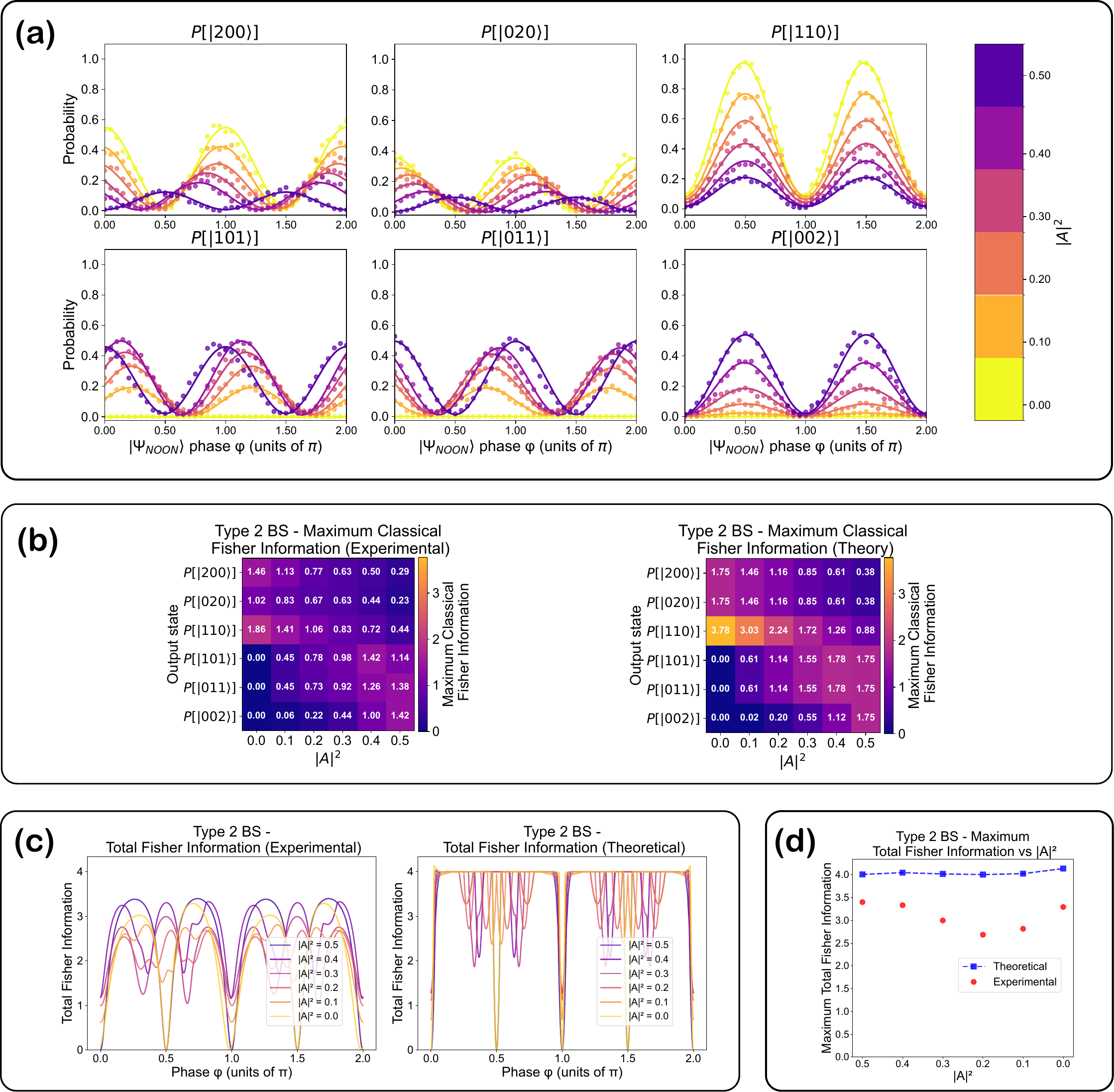}
\caption{\textbf{Fisher information analysis for the NOON state experiment (Type~2 configuration).}
}
\label{figS9}
\end{figure}

\newpage


\end{document}